\documentclass[aps,prl,twocolumn,superscriptaddress]{revtex4-2}

\usepackage{graphicx}
\usepackage{dcolumn}
\usepackage{bm}
\usepackage{rotating}

\usepackage{bm,color}

\usepackage{tabularx}
\usepackage{epstopdf}

\usepackage{ulem} 
\usepackage{hyperref}

\usepackage{tikz-feynman}

\usepackage{overpic}
\usepackage{verbatim}

\usepackage{lineno}

\newcommand{\jpsi}{J/\psi}
\newcommand{\EE}{e^+e^-}
\newcommand{\MM}{\mu^+\mu^-}

\newcommand{\lum}{{\cal L}}

\newcommand{\cco}{\chi_{c1}}

\newcommand{\xx}{X(3872)}
\newcommand{\ev}{\mathrm{eV}}

\newcommand{\mev}{\mathrm{MeV}}

\newcommand{\gev}{\mathrm{GeV}}

\newcommand{\ivpb}{\mathrm{pb}^{-1}}

\newcommand{\gammaee}{\Gamma_{ee}}

\begin{document}

\title{\boldmath First Observation of the Direct Production of the $\cco$ in $\EE$ Annihilation}

\author{
\begin{small}
\begin{center}
M.~Ablikim$^{1}$, M.~N.~Achasov$^{10,b}$, P.~Adlarson$^{69}$, S. ~Ahmed$^{15}$, M.~Albrecht$^{4}$, R.~Aliberti$^{29}$, A.~Amoroso$^{68A,68C}$, M.~R.~An$^{33}$, Q.~An$^{65,51}$, X.~H.~Bai$^{59}$
% ...
\\
\vspace{0.2cm}
(BESIII Collaboration)\\
\vspace{0.2cm} {\it
$^{1}$ Institute of High Energy Physics, Beijing 100049, People's Republic of China\\
% ...
$^{4}$ Bochum Ruhr-University, D-44780 Bochum, Germany\\
% ...
$^{10}$ G.I. Budker Institute of Nuclear Physics SB RAS (BINP), Novosibirsk 630090, Russia\\
% ...
$^{15}$ Helmholtz Institute Mainz, Staudinger Weg 18, D-55099 Mainz, Germany\\
% ...
$^{29}$ Johannes Gutenberg University of Mainz, Johann-Joachim-Becher-Weg 45, D-55099 Mainz, Germany\\
% ...
$^{33}$ Liaoning Normal University, Dalian 116029, People's Republic of China\\
% ...
$^{51}$ State Key Laboratory of Particle Detection and Electronics, Beijing 100049, Hefei 230026, People's Republic of China\\
% ...
$^{59}$ University of Jinan, Jinan 250022, People's Republic of China\\
% ...
$^{65}$ University of Science and Technology of China, Hefei 230026, People's Republic of China\\
% ...
$^{68}$ University of Turin and INFN, (A)University of Turin, I-10125, Turin, Italy; (B)University of Eastern Piedmont, I-15121, Alessandria, Italy; (C)INFN, I-10125, Turin, Italy\\
$^{69}$ Uppsala University, Box 516, SE-75120 Uppsala, Sweden\\
% ...
$^{b}$ Also at the Novosibirsk State University, Novosibirsk, 630090, Russia.\\
% ...
}\end{center}
\vspace{0.4cm}
\end{small}
}

\begin{abstract}
We study the direct production of the $J^{PC}=1^{++}$ charmonium state 
$\cco(1P)$ in electron-positron annihilation by carrying out an energy 
scan around the mass of the $\cco(1P)$. The data were
collected with the BESIII detector at the BEPCII collider.
An interference pattern between the signal process
$\EE\to \cco(1P) \to \gamma\jpsi\to\gamma\MM$
and the background processes
$\EE\to \gamma_{\mathrm{ISR}}\jpsi\to\gamma_{\rm{ISR}}\MM$
and $\EE \to\gamma_{\rm{ISR}}\MM$
are observed by combining all the data samples.
The $\cco(1P)$ signal is observed with a significance of 5.1$\sigma$.
This is the first observation of a $C$-even state directly produced in
$\EE$ annihilation.
The electronic width of the $\cco(1P)$ resonance is determined to be
$\gammaee=(0.12^{+0.13}_{-0.08})~\ev$,
which is of the same order of magnitude as theoretical calculations.
\end{abstract}

\maketitle

In the process $\EE\to R$,  where $R$ represents a hadronic resonance,
the dominant production mechanism, when allowed,
is through one virtual photon. This results in the copious production 
of vector mesons with $J^{PC}=1^{--}$, where the quantum numbers $J$, 
$P$, and $C$ denote the spin,
parity, and charge conjugation of $R$, respectively.
In principle, $C$-even resonances
can also be produced directly in $\EE$ annihilation
through processes with two timelike virtual photons or neutral currents.
Notice that the production via two real photons is forbidden
due to the Landau-Yang theorem.
Such processes were discussed already  
40 years ago~\cite{40 years discuss} and 
were revisited in 
Refs.~\cite{Yang:2012gk,Achasov:1996gt,Achim-predict-X3872,Marc-gee-chic1}.
Experimental searches for $\EE$ annihilation to the $\eta$, $\eta'$, $f_{0}(980)$, $f_{0}(1300)$,
$f_{1}(1285)$, $f_{2}(1270)$, $a_{0}(980)$, $a_{2}(1320)$,
and the $\xx$ [also known as $\cco(3872)$] have been
carried out at the VEPP-2M~\cite{ND-detector,ee-to-a2(1320)-and-f2(1270),ee-to-eta},
VEPP-2000~\cite{ee-to-eta'-CMD-3,ee-to-eta',ee-to-f1(1285)},
and BEPCII~\cite{ee-to-X3872} colliders.
The most significant signal (2.5$\sigma$) 
was obtained for the $f_{1}(1285)$~\cite{ee-to-f1(1285)}.
All others resulted in upper limits on the electronic
widths ($\gammaee$) of the corresponding resonances.
In a spacelike two-photon scattering process, $\EE\to\EE\xx$, evidence ($3.2\sigma$) for the 
$\xx$ production has also been found~\cite{gg-to-X3872-at-belle} at Belle. 
As for the $\cco(1P)$, which we refer to as the $\cco$,
there have been no previous searches.

\begin{figure}[htbp]
\begin{center}
\includegraphics[width=0.45\textwidth]{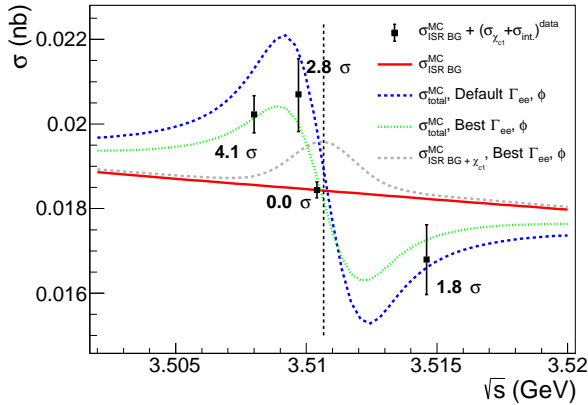}
\caption{
The colored curves are energy-dependent cross sections of the process
$\EE\to\gamma\jpsi\to\gamma\MM$ including (green and blue curves) and not including (red curve) the direct
production of $\EE\to\cco$ (see text for more details).
The gray curve denotes the signal strength in the hypothetical case of no interference. 
The location of the $\cco$ mass is indicated by the vertical line. 
The black dots with
error bars represent $\sigma_{\rm ISR~BG}^{\rm MC} + (\sigma_{\cco}+\sigma_{\rm int})^{\rm data}$ at the $\cco$ scan data samples. The numbers next to the four data points indicate the statistical significances associated with the $\cco$ production.
}
\label{fig:chic1 scan sample}
\end{center}
\end{figure}

Following the strategy for calculating the electronic width 
of the $\cco$ suggested in Ref.~\cite{40 years discuss},
the authors of Ref.~\cite{Henryk-production} predict $\gammaee=0.41~\ev$.
This work also considers the interference
between the signal process, $\EE\to\cco\to\gamma\jpsi\to\gamma\MM$,
and the irreducible background processes
$\EE\to\gamma_{\rm ISR}\jpsi\to\gamma_{\rm ISR}\MM$ and nonresonant
$\EE\to\gamma\MM$, see blue and red curves in Fig.~\ref{fig:chic1 scan sample}. Here ISR stands for initial state radiation. 
Depending on the value of the relative
phase $\phi$ between the signal and background amplitudes, the interference
changes the total cross section line shape dramatically.

In this Letter, we report a search for the reaction $\EE\to\cco$ at the BESIII experiment at the BEPCII collider.
First, the background processes 
are studied and then we carry out a search 
for the signal process beyond the background.
The data samples are collected at four center-of-mass (c.m.)
energies (3.5080, 3.5097, 3.5104, and
3.5146 GeV) in the $\cco$ mass region 
(referred to as the $\cco$ scan sample)
with the BESIII detector~\cite{Ablikim:2009aa}.
The first two scan points are located below the $\cco$ mass, 
where according to Ref.~\cite{Henryk-production} a constructive interference effect between the signal 
process and the irreducible background processes
is expected.
The third scan point is very slightly below the mass position,
hence a minimal effect is predicted. The fourth point is above 
the $\cco$ mass, which should lead to a reduction of events with 
respect to the scenario with no direct production of the $\cco$.
If there was no interference, the excess at the third point would be
expected to be the largest (see gray line in Fig.~\ref{fig:chic1 scan sample}).
The data samples are listed in Table~\ref{table:data sample}.
The c.m. energies are
measured using a beam energy measurement system (BEMS)~\cite{BEMS-detector} with an uncertainty of $\pm0.05~\mev$ 
and the beam-energy spread is measured to be $(736\pm27)$ keV. 
The total integrated luminosity of the four data samples
is $446~\ivpb$, which is measured using large angle Bhabha events. 
To verify the background description,
we have also analyzed four already existing control samples, 
in which the signal process is absent,
with a total integrated luminosity of $6294~\ivpb$,
of which two samples have a large integrated luminosity
($\sqrt{s}=3.773~\gev$, and $4.178~\gev$), while the other two
are comparable in size to the scan samples ($\sqrt{s}=3.581~\gev$ and $3.670~\gev$),
as summarized in Table~\ref{table:data sample}.

The $\cco$ is reconstructed via its radiative decay
$\cco\to\gamma\jpsi$, with
the subsequent decay $\jpsi\to\MM$.
The $\jpsi\to\EE$ mode is not used due
to large background from the Bhabha process~($\EE\to\EE$).

\begin{table*}[htbp]
\caption{
The c.m. energies, integrated luminosities, and fit results
for the control samples
(above the horizontal line) and for the $\cco$ scan sample (below).
The number of signal events~($N_{\rm sig}$) is obtained from a two-dimensional
fit without ($N_{\rm sig}$ w/o Corr.) and with ($N_{\rm sig}$ w/ Corr.)
the two-dimensional correction described in the text.
The first uncertainty is statistical, and the second is systematic (if applied).
The first value in parentheses denotes the statistical significance.
The second value in parentheses for the control samples is the significance when the size of the data set is normalized to $180~\ivpb$ (thereby increasing the statistical errors); the second value for the
$\cco$ scan samples is the significance
including the systematic
uncertainties.
 The last column shows
the number of signal events
derived from a MC sample where the values of $\gammaee$ and $\phi$
are fixed to the values obtained from
a common fit to all $\cco$ scan samples
(the error includes systematic effect).}
\label{table:data sample}

\begin{center}
\begin{tabular}{lrrrc}
\hline\hline
$\sqrt{s}$ ($\mev$) &~~$\lum$ ($\ivpb$)~~& $N_{\rm sig}$ w/o Corr. ~~~~~~~~
& $N_{\rm sig}$ w/ Corr. ~~~~~~~~ & $N_{\rm sig}$ w/ Corr. common fit
\\\hline\hline

$3773.0$&  $2932.4~~~$            
&$1027\pm 140$~($7.5\sigma$;~$1.9\sigma_{180}$)
& ~$49\pm 141$~($0.3\sigma$;~$0.1\sigma_{180}$)~~~~~
&   $\cdots$ \\

$4178.4$&  $3192.5~~~$            
&$522 \pm 104$~($5.1\sigma$;~$1.2\sigma_{180}$)
& ~$40\pm 104$~($0.4\sigma$;~$0.1\sigma_{180}$)~~~~~
&  $\cdots$  \\
$3581.5$&  $85.3~~~$              
&$31\pm 29$~($1.1\sigma$;~$1.6\sigma_{180}$)~
& $-5\pm 29$~($0.2\sigma$;~$0.3\sigma_{180}$)~~~~~~
&   $\cdots$  \\
$3670.2$&  $83.6~~~$              
&$38\pm 26$~($1.5\sigma$;~$2.2\sigma_{180}$)~
& ~~$4\pm 26$~($0.2\sigma$;~$0.2\sigma_{180}$)~~~~~~
&  $\cdots$ \\

\hline

$3508.0$ &  $181.8~~~$
& $320\pm 51$~($6.5\sigma$)~~~~~~~~~~~~ 
& $210\pm 52\pm 18$~($4.1\sigma$;~$4.0\sigma_{\rm low}$) 
&
$191^{+60}_{-59}$
\\
$3509.7$ &  $39.3~~~$
& $85\pm 24$~($3.9\sigma$)~~~~~~~~~~~~ 
& $63\pm 24\pm 6$~~($2.8\sigma$;~$2.7\sigma_{\rm low}$) 
&  ~~$41^{+20}_{-19}$
\\
$3510.4$ &  $183.6~~~$
& $100\pm 48$~($1.7\sigma$)~~~~~~~~~~~~ 
& ~~~~~~$ 0^{+16}_{-19}\pm 23$~~($0.0\sigma$;~$0.0\sigma_{\rm low}$)
& ~~$42^{+79}_{-77}$
\\
$3514.6$ &  $40.9~~~$
& ~~$-16^{+16}_{-21}$~($0.7\sigma$)~~~~~~~~~~~~~ 
& $-40\pm 22 \pm 7$~($1.8\sigma$;~$1.6\sigma_{\rm low}$)  
&  $-29^{+8}_{-10}$~
\\

Combined &  $445.6~~~$   & $\cdots$~~~~~~~~~~~~~~~~~~~~  
& $\cdots$~($5.3\sigma$;~$5.1\sigma_{\rm low}$)~~~~~~
& $\cdots$~($5.1\sigma$;~$4.2\sigma_{\rm low}$)\\
\hline\hline
\end{tabular}
\end{center}
\end{table*}

Monte Carlo (MC) samples are used to determine the detection efficiencies
and to estimate the background contributions.
Simulated samples are produced with a {\sc geant4}-based~\cite{geant4} MC package, which includes the geometric description of
the BESIII detector and the detector response.
The \textsc{phokhara}~\cite{PHOKHARA_web} event generator is used to describe
the signal process
($\EE\to\cco\to\gamma\jpsi\to\gamma\MM$),
the irreducible background processes
($\EE\to\gamma_{\rm ISR}\jpsi\to\gamma_{\rm ISR}\MM$ and
$\EE\to\gamma\MM$),
and the interference between them.
Angular distributions for the signal process are implemented
into the \textsc{phokhara} event generator
using Ref.~\cite{Henryk-production},
while the background ISR processes are modeled using
Ref.~\cite{PHOKHARA_web}.
Non-$\gamma_{(\rm ISR)}\MM$ background events are 
found to be negligible ($<0.2\%$) by studying
control samples~\cite{BESIII:2021yal gammaChicj} and
inclusive MC simulations,
which include the production of open-charm mesons, the ISR
production of vector charmonium(like) states,
and continuum processes.

A full reconstruction method is used to select
$\gamma\MM$ candidate events.
The charged tracks and photons are selected with the same method
as described in Ref.~\cite{BESIII:2021yal gammaChicj}.
Muon tracks are identified by
the energy they deposit in the electromagnetic calorimeter
(EMC) and requiring $E_{\rm EMC}<0.4~\gev$.
A four constraint (4$C$) kinematic fit is applied with two charged
tracks and one of the photons constraining the total reconstructed
four momentum to that of the initial state. The photon with minimum
$\chi^{2}_{4C}$ is chosen as the best photon candidate. 
As checked within a MC simulation, the probability to select a wrong photon is negligible.
We require the polar angle of the best
photon candidate to be $|\cos\theta_{\gamma}|<0.80$ to suppress
background events from ISR processes.

The verification of background description
is done quantitatively by performing a two-dimensional
fit to the $\MM$ invariant mass ($M_{\MM}$) distribution
and the $|\cos\theta_{\mu}|$ distribution
with noninterfering signal and background components,
whose line shapes are extracted from the corresponding MC simulations.
The signal line shape is taken from the $\cco$ signal MC simulation at $\sqrt{s}=3.5080~$GeV,
smeared with two Gaussian functions, one
to account for the resolution difference between data and MC
simulations
and the other for line-shape differences between different energy points.
In the $\MM$ invariant mass distribution, we expect the irreducible
background events to feature a $\jpsi$ peak
($\EE\to \gamma_{\mathrm{ISR}}\jpsi\to\gamma_{\rm{ISR}}\MM$)
on top of a smooth distribution ($\EE \to\gamma\MM$).
The relative sizes of these background
contributions are fixed
using our best estimate for the electronic
width of the $J/\psi$ ($\Gamma_{ee}^{J/\psi}$).
The number of signal events ($N_{\rm sig}$)
is expected to be
zero in the control samples.
The statistical significance of the signal contribution
is determined by the difference of the best
log-likelihood ($-\ln{L}$) value and
the log-likelihood value for a fit with null-signal hypothesis.
However, as summarized in the third column of Table~\ref{table:data sample}, nonzero values for
$N_{\rm sig}$ have been found,
representing a discrepancy between the data and the
MC simulation of the irreducible background process. 
We have verified that this
discrepancy is not due to differences between data and MC simulation in the experimental efficiencies,
but rather can be explained by
uncertainties in the input
$\Gamma_{ee}^{J/\psi}$
and
limitations of the {\sc phokhara} event generator
in simulating the ISR production of the narrow $\jpsi$ resonance for large-angle
ISR photons~\cite{CarloniCalame:2011zq}.
The statistical significance of the
discrepancy differs sizably for the four control samples. When normalizing
the effect of the discrepancy to an integrated luminosity of $180~\ivpb$,
which corresponds to a typical luminosity of the $\cco$ scan points, we observe
significances below 2.3$\sigma$.

We carry out a two-dimensional correction to the distributions of $M_{\MM}$ and
$|\cos\theta_{\mu}|$
by re-weighting MC simulated events
to correct the discrepancy.
The correction factors are extracted using data and MC samples at $\sqrt{s}=3.773~\gev$
or $4.178$~GeV and are applied to the
MC simulations at other data samples (see
Supplemental Material~\cite{ref. supplemental material}).
After applying these correction factors, $N_{\rm sig}$ is consistent with
zero within one standard deviation for all control samples.

In order to extract the number of signal events
at the four $\cco$ scan points,
the $M_{\MM}$ and $|\cos\theta_{\mu}|$ distributions are investigated using a similar method as above.
The fit is performed at each data sample individually
using a two-dimensional unbinned maximum likelihood fit method.
The line shapes for the contributions from the $\cco$ production,
the irreducible background,
and the interference between them are derived from the corresponding
individual MC simulations (see Supplemental Material~\cite{ref. supplemental material} for the angular distributions).
The same two-dimensional correction as above is applied to the shapes of the background processes,
and the square root of the same factor is used for
the interference.
The numbers of $\cco$
($N_{\cco}$) and 
irreducible
background events ($N_{\rm bg}$) are free parameters, while the
interference ($N_{\rm int}$) is written as $f \cdot \sqrt{N_{\cco}\cdot N_{\rm bg}}$,
where the factor $f$ is determined from signal MC sample
with the $\gammaee$ and $\phi$
parameters set to the optimal values from a common fit to all scan points, as
will be explained below.

\begin{figure}[htbp]
\begin{center}
\includegraphics[width=0.48\textwidth]{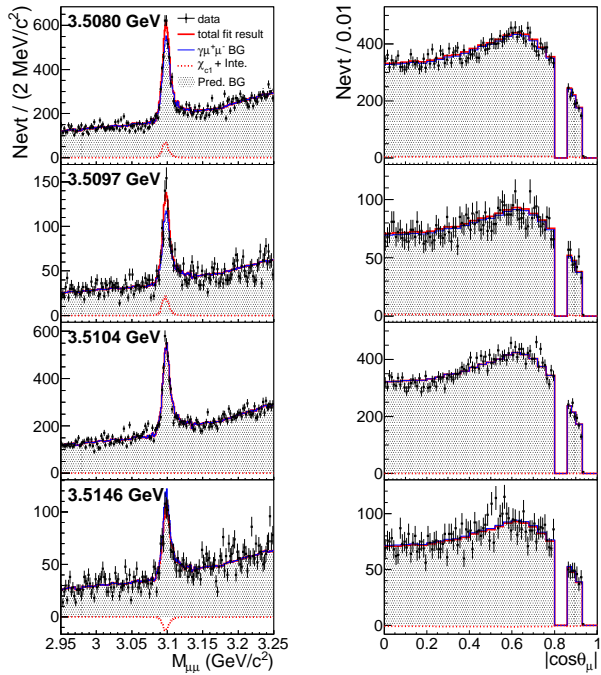}
\end{center}
\caption{One-dimensional projections of the two-dimensional fit
to the $M_{\MM}$ and $|\cos\theta_{\mu}|$ distributions from the $\cco$ scan samples.
The black dots with error bars are 
from data, the gray histograms
are the irreducible background predicted by the corrected MC simulation.
The red curve is the
best fit result, the red dotted (blue dashed) curve is the signal (background)
contribution.
The region between $0.8<|\cos\theta_{\mu}|<0.86$ corresponds to 
the gap between the barrel and end cap modules of the EMC.
}
\label{fig:gmm_fit_result_2d}
\end{figure}

The fit results
are shown in Fig.~\ref{fig:gmm_fit_result_2d}
and are listed in Table~\ref{table:data sample}.
Significant signal components are
seen at $\sqrt{s}=3.5080~\gev$ and $3.5097~\gev$ with
$N_{\rm sig} = N_{\chi_{c1}} + N_{\rm int}$ (and its statistical significance)
determined to be $210\pm 52$ (4.1$\sigma$)
and $63\pm 24$ (2.8$\sigma$), respectively.
The signal component is not significant at $\sqrt{s}=3.5104~\gev$ with
$N_{\rm sig} = 0^{+16}_{-19}$ (0.0$\sigma$).
A negative signal component is seen at $\sqrt{s}=3.5146~\gev$
with $N_{\rm sig}=-40\pm 22$ (1.8$\sigma$).
The combined statistical significance,
obtained by adding the
log-likelihoods
from each of the
four data
samples, is $5.3\sigma$.
The cross section of the signal component and its uncertainty is calculated
as $\sigma_{\rm sig} \equiv  (\sigma_{\cco}+\sigma_{\rm int})^{\rm data} = N_{\rm sig}/(\lum\cdot\epsilon)$,
where the efficiency
$\epsilon$ is calculated from the simulated signal MC samples. The sum of
$\sigma_{\rm sig}$ and $\sigma_{\rm ISR~BG}$
at each $\cco$ scan point is
shown in Fig.~\ref{fig:chic1 scan sample} (black dots), which is in good
agreement with the theoretical prediction~\cite{Henryk-production}. Here
$\sigma_{\rm ISR~BG}$ is fixed using the \textsc{phokhara} generator.
Statistical tests are performed to the $\cco$ scan samples individually
using likelihood ratios
$t=-(\ln{L_{\rm s}}-\ln{L_{\rm ns}})$ to discriminate
the hypothesis with or without 
signal components (distributions to be found in
Supplemental Material~\cite{ref. supplemental material}).

\begin{figure}[htbp]
\begin{center}
\includegraphics[width=0.45\textwidth]{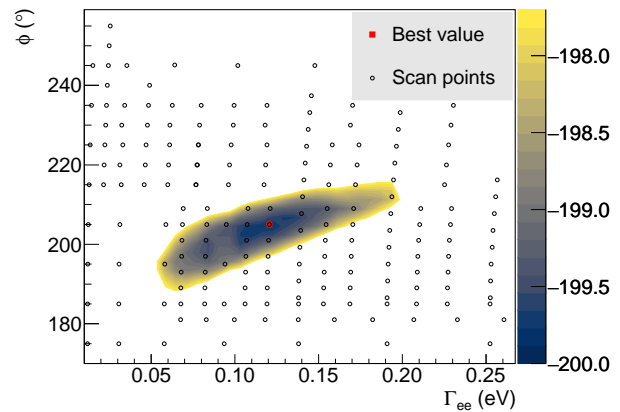}
\end{center}
\caption{
The 68.3\% C.L. contour of $\gammaee$ and $\phi$
on a distribution of log-likelihood ($-\ln{L}$) values.
The distribution of $-\ln{L}$ in a larger parameter space region
is shown in Supplemental Material \cite{ref. supplemental material}.
}
\label{fig:gmm_chi2_scan_result}
\end{figure}

Using a common fit to the four $\cco$ scan points, the values of $\gammaee$
and $\phi$ can be determined directly from data. Since it is not easy to obtain
an analytic formula for the total cross section of $\EE\to\gamma_{(\rm ISR)}\MM$
as a function of $\gammaee$ and $\phi$, the analysis is done via a scan method.
At each c.m. energy of the $\cco$ scan sample, the MC samples of
$\EE\to\gamma_{(\rm ISR)}\MM$ are produced with different sets of
($\gammaee$, $\phi$) values, see open circles in Fig.~\ref{fig:gmm_chi2_scan_result}.
The total likelihood from the
four samples in the $\cco$ mass region
is then calculated
using the same two-dimensional distributions used previously
with the number of events
at each energy point
constrained
to the expected number of events calculated from MC.
The best $\gammaee$ and $\phi$ parameters are determined to be
$(0.12^{+0.08}_{-0.07})$ eV and $(205.0^{+10.0}_{-17.0})^{\circ}$, respectively,
where the uncertainty corresponding to $68.3\%$ C.L.
is statistical only. The $68.3\%$ C.L. contour region in
the ($\gammaee$, $\phi$) plane is shown in Fig.~\ref{fig:gmm_chi2_scan_result},
in which the red dot represents the best-fitted value.
The green curve in Fig.~\ref{fig:chic1 scan sample} shows the 
cross section line shape for such a 
set of parameters. 
Using this best set of
($\gammaee, \phi$) values,
the number of signal events is estimated for each
$\cco$ scan sample and is found to be 
191, 41, 42, $-$29
events for the four scan samples.
The uncertainties on
$N_{\rm sig}$
are estimated by varying
the ($\gammaee, \phi$) values within their $68.3\%$ C.L. contour and
finding the largest variations of $N_{\rm sig}$.
Combining the four samples, 
the statistical significance is 5.1$\sigma$ and is
found to be in very good agreement with the previous estimate by fitting each scan sample individually, where
$\gammaee$ and
$\phi$ are not constrained to be the same.

Systematic uncertainties
for the extraction of $\gammaee$ and $\phi$
mainly come from the luminosity measurement, 
the detection efficiency, the
line shapes used in the fit, the fit range, the two-dimensional correction
factor, the non-$\gamma_{\rm (ISR)}\MM$ background contribution, and the c.m.
energy measurement.

The systematic uncertainty on
the measurement of the integrated luminosity
is $0.6\%$ for each data sample.
We take 0.5\% as the uncertainty for muon reconstruction,
which is assumed to be the same as for electron reconstruction.
The uncertainty in photon reconstruction is
estimated to be $0.2\%$, obtained using control samples of the $\EE\to\gamma\MM$ process.
The systematic uncertainties
from the integrated luminosity measurement and
detection efficiency are considered simultaneously
by changing the normalization factor
used in the scan fit by $1.0\%$.
The uncertainty from 
the requirement on $|\cos\theta_{\gamma}|$
is studied by tightening the requirement
from 0.8 to 0.79,
0.78, 0.77, and 0.76, the largest deviation with respect to the
default one is taken as the systematic uncertainty.
Systematic uncertainties from other selection criteria are negligible.

The uncertainties from the binning strategy and the fit procedure
are studied using toy MC samples,
no bias is found.
The uncertainty from the beam energy spread is considered by changing it
from 736 to 1000 keV, the change is much larger than its standard deviation
measured by BEMS (27 keV).
The fit range of the $M_{\MM}$ distribution is varied and the difference between the nominal result is considered as the systematic uncertainty.
The uncertainty from the two-dimensional correction factor is estimated
by replacing the nominal one extracted from the $\sqrt{s}=3.773~\gev$ data sample
with that from the $\sqrt{s}=4.178~\gev$ data sample.
In addition, the square root of the correction factor
is applied to the interference term based on the assumption that the discrepancy
observed at the control sample comes entirely
from the generator level.
The uncertainty from this assumption is studied by dropping
the correction to the interference term.
The non-$\gamma_{\rm (ISR)}\MM$ background contribution
is neglected in the nominal fit,
the uncertainty from it is considered by including it.

We change the $\sqrt{s}$ in MC simulation at each energy point by
$\pm0.05~\mev$ and take the changes as systematic uncertainty from the c.m. energy.
Assuming all the systematic uncertainties are uncorrelated and adding
them in quadrature, the largest parameter ranges of $\gammaee$ and $\phi$
corresponding to 68.3\% C.L.
are determined to be $(0.12^{+0.13}_{-0.08})~\ev$
and $(205.0^{+15.4}_{-22.4})^{\circ}$, respectively.
The total systematic uncertainties are of a similar size as the
statistical effects.
After having estimated the statistical and systematic uncertainties
associated with our fit to $\gammaee$ and $\phi$, we study the
dependence of signal events
by varying these input parameters within the contour determined at
68.3\% C.L., as listed
in the last column of Table~\ref{table:data sample}.

Systematic uncertainties for the individual fits are estimated using similar
methods as listed above. However, when considering the systematic uncertainties on $N_{\rm sig}$, the one on the requirement
of $|\cos\theta_{\gamma}|$ is excluded since the signal yields change.
One extra term comes from the input $\gammaee$ and $\phi$ values, which
affect the signal line shape and is considered by varying the values
within the 68.3\% C.L. contour.

As summarized in Table~\ref{table:data sample}, we list the minimum significance found both in the case of
individual fits (column
``$N_{\rm sig}$ w/ Corr.'')
and in the case of a common fit (column ``$N_{\rm sig}$ w/ Corr. common fit'').
After including the systematic uncertainties, the
minimum significance is found to be 5.1$\sigma$ in the first and 4.2$\sigma$ in the second case.
As the significance obtained by combining individual fits is more robust 
to systematic effects and does not rely on the specific model of Ref.~\cite{Henryk-production},
we take it as our nominal result.

In summary, using data samples taken
in the $\cco$ mass
region, we observe the direct production of the $C$-even resonance, $\cco$,
in $\EE$ annihilation for the first time with a statistical significance
larger than 5$\sigma$. We observe a typical interference pattern around
the $\cco$ mass, which previously
was predicted in
Ref.~\cite{Henryk-production}.
The electronic width of the $\cco$ has been determined
for the first time from a common fit to the four scan samples to be
$\gammaee=(0.12^{+0.13}_{-0.08})~\ev$.
This observation demonstrates that with the current generation of
electron-positron colliders, the direct production of $C$-even resonances
through two virtual photons is possible. As 
a next step, we intend to embark on a scan around the $\chi_{c2}$ resonance at BESIII.
Using future super-tau-charm factories with increased luminosity~\cite{Lyu:2021tlb},
the $\gammaee$ and other properties such as the line shapes of $C$-even
states could be determined by performing a similar scan method.
This will shed light on the intrinsic nature of charmoniumlike resonances.

The BESIII Collaboration thanks the staff of BEPCII and the IHEP computing center for their strong support.
We acknowledge the continuous theory support for this analysis by
Henryk Czy{\.z} and Hans K{\"u}hn.
This work is supported in part by National Key R\&D Program of China under Contracts No.
2020YFA0406300, No. 2020YFA0406400; 
National Natural Science Foundation of China (NSFC) under Contracts 
No. 11635010, No. 11735014, No. 11835012, No. 11935015, No. 11935016, No. 11935018, No. 11961141012, No. 12022510, No. 12025502,
No. 12035009, No. 12035013, No. 12192260, No. 12192261, No. 12192262, No. 12192263, No. 12192264,
No. 12192265; 
the Chinese Academy of Sciences (CAS) Large-Scale Scientific Facility Program; 
Joint Large-Scale Scientific Facility Funds of the NSFC and CAS under Contracts No. U1832207, No. U2032108; 
CAS Key Research Program of Frontier Sciences under Contract No.
QYZDJ-SSW-SLH040; 100 Talents Program of CAS; 
Shanghai Pujiang Program (20PJ1401700);
INPAC and Shanghai Key Laboratory for Particle Physics and Cosmology; 
ERC under Contract No. 758462; 
European Union's Horizon 2020 research and innovation programme under Marie Sklodowska-Curie grant agreement under Contract No. 894790; 
German Research Foundation DFG under Contracts Nos. 443159800, 
Collaborative Research Center CRC 1044, GRK 2149; 
Istituto Nazionale di Fisica Nucleare, Italy; 
Ministry of Development of Turkey under Contract No. DPT2006K-120470; 
National Science and Technology fund; 
STFC (United Kingdom); 
The Royal Society, UK under Contracts No. DH140054, No. DH160214; 
The Swedish Research Council;
U.~S. Department of Energy under Contract 
No. DE-FG02-05ER41374.

\bibliography{basename of .bib file}

\end{document}


\title{\boldmath Supplemental Material for ``First observation of the direct production of the $\cco$ in $\EE$ annihilation"}

\author{
BESIII Collaboration
}

\maketitle

\onecolumngrid

\section{The distribution of -ln(L) in a larger parameter space region}

\begin{figure}[htbp]
\begin{center}
\includegraphics[width=0.45\textwidth]{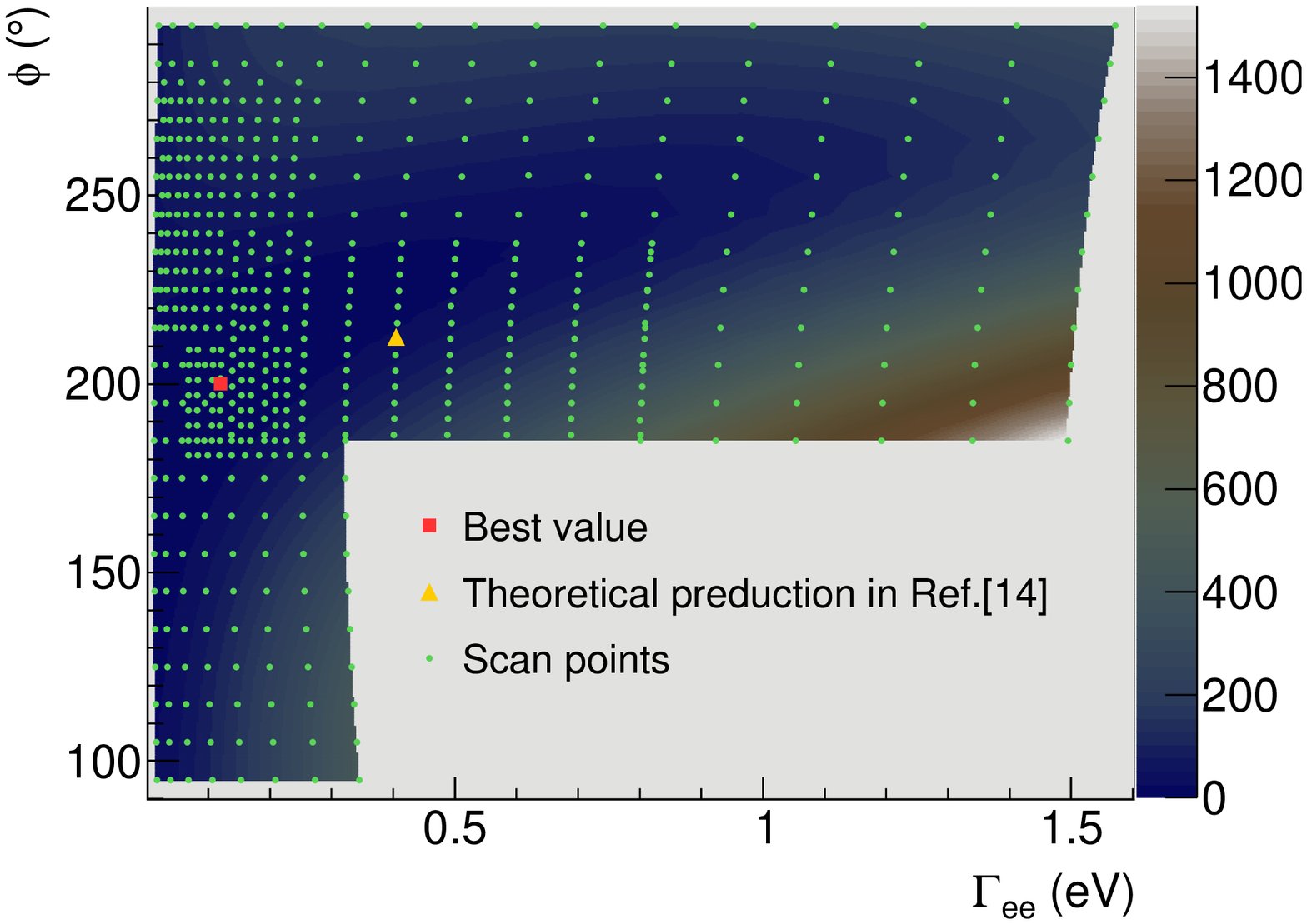}
\end{center}
\caption{
The distribution of $-\ln(L)$ in a larger parameter space region.
}
\label{fig:gmm_chi2_scan_result2}
\end{figure}

Figure~\ref{fig:gmm_chi2_scan_result2} shows the distribution of the log-likelihood 
value ($-\ln(L)$) as a function of $\Gamma_{ee}$ ($x$-axis) and $\phi$ ($y$-axis) in 
a larger parameter space region. The red square ($0.12~\mathrm{eV}$, $205.0^{\circ}$) 
represents the point where the likelihood value is maximum. The orange triangle 
($0.41~\mathrm{eV}$, $212.0^{\circ}$) comes from the theoretical calculation in 
Ref.~[14]. The
green circles are the parameter points where MC samples are produced.

\section{The correction factor}

\begin{figure}[htb]
\begin{center}
\includegraphics[width=0.45\textwidth]{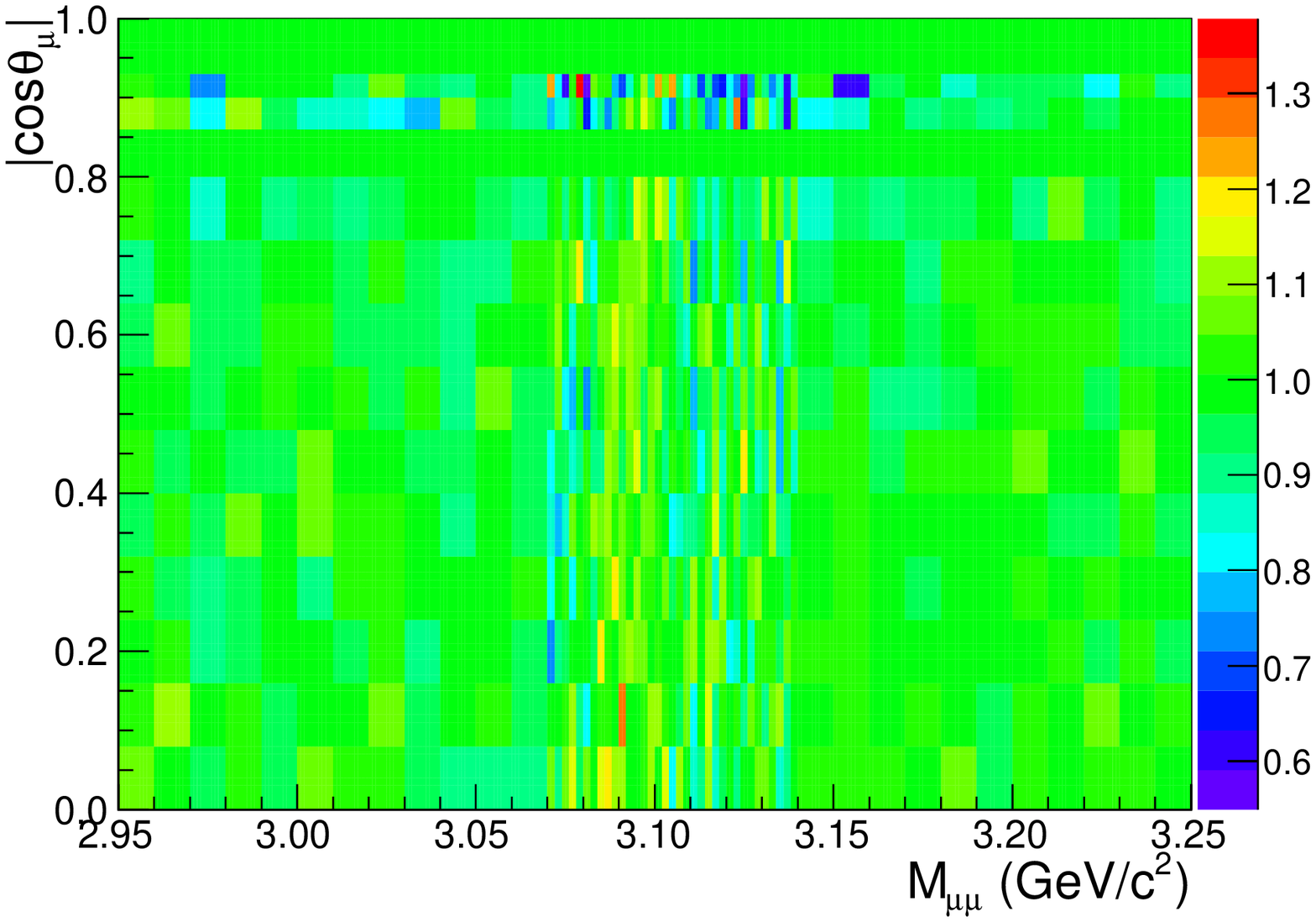}
\includegraphics[width=0.45\textwidth]{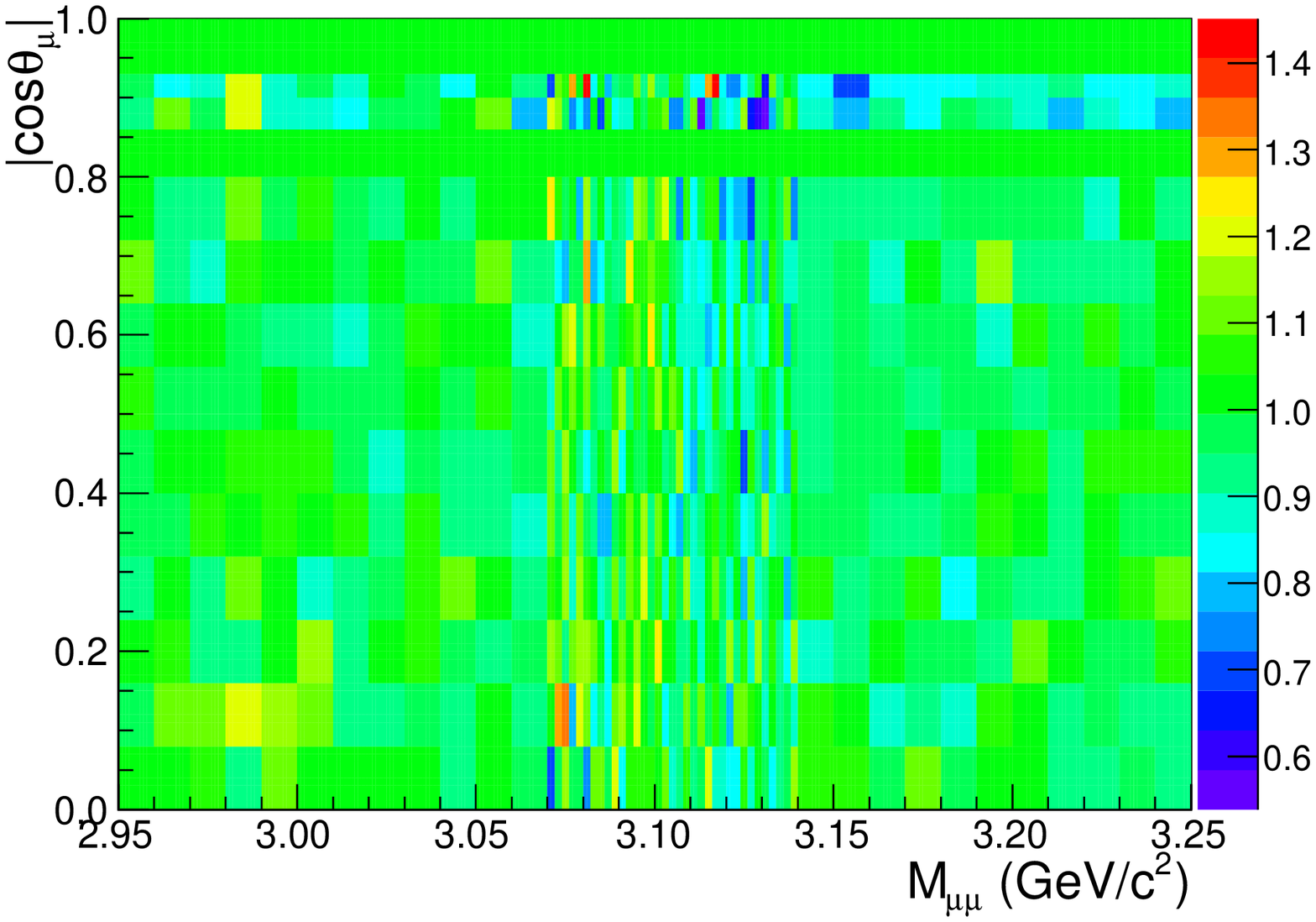}
\end{center}
\caption{The correction factors extracted from the $\sqrt{s}=3.773~\mathrm{GeV}$ sample
(left) and the $\sqrt{s}=4.178~\mathrm{GeV}$ sample (right).
}
\label{fig:corr_factor}
\end{figure}

\begin{figure*}[htb]
\begin{center}
\includegraphics[width=0.45\textwidth]
{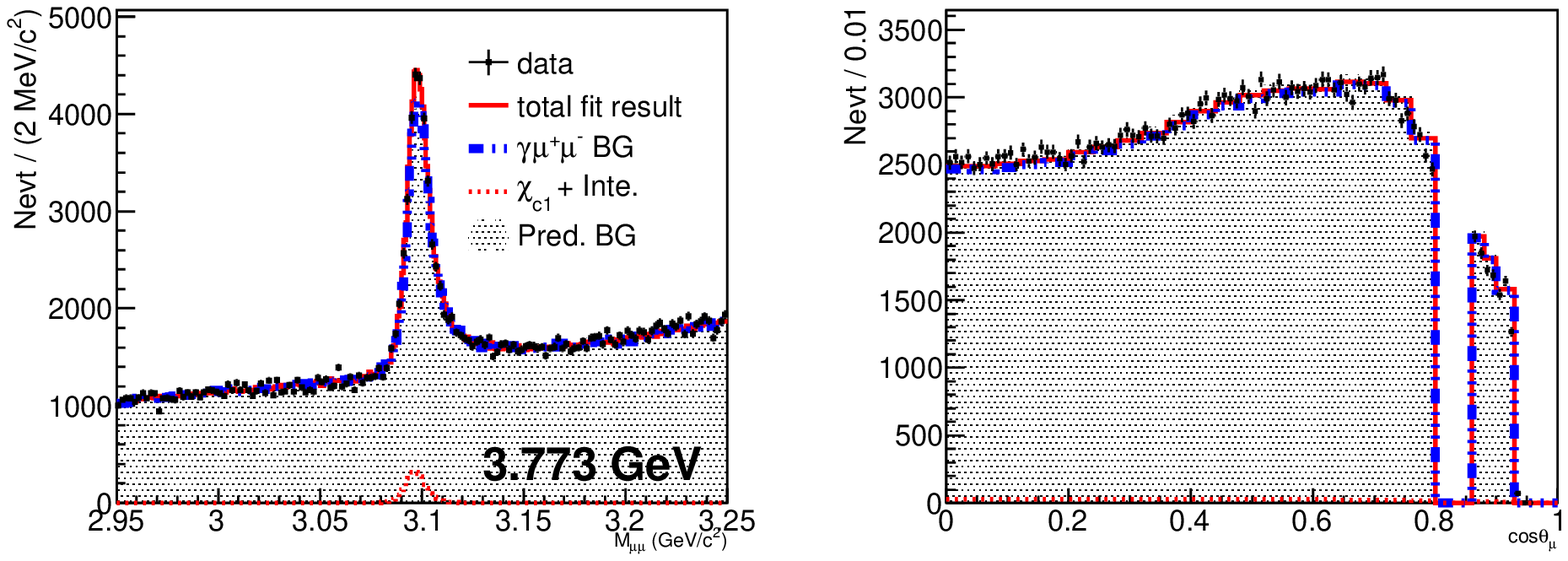}
\includegraphics[width=0.45\textwidth]
{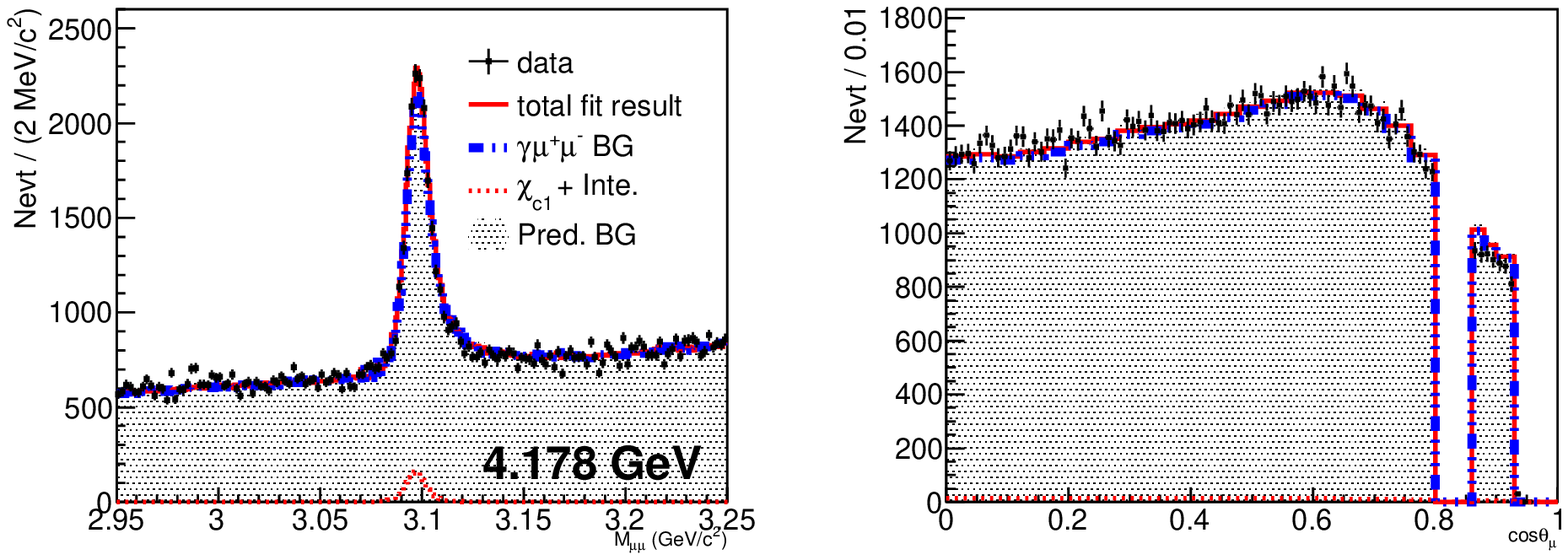}
\includegraphics[width=0.45\textwidth]
{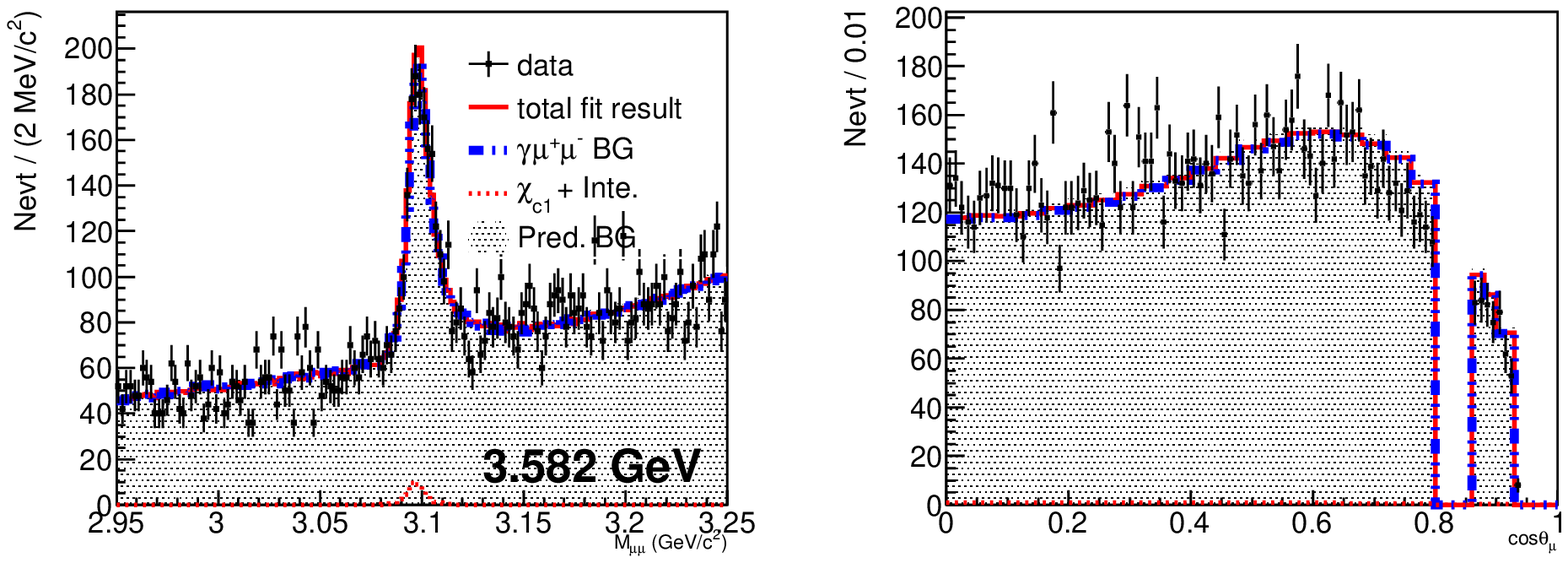}
\includegraphics[width=0.45\textwidth]
{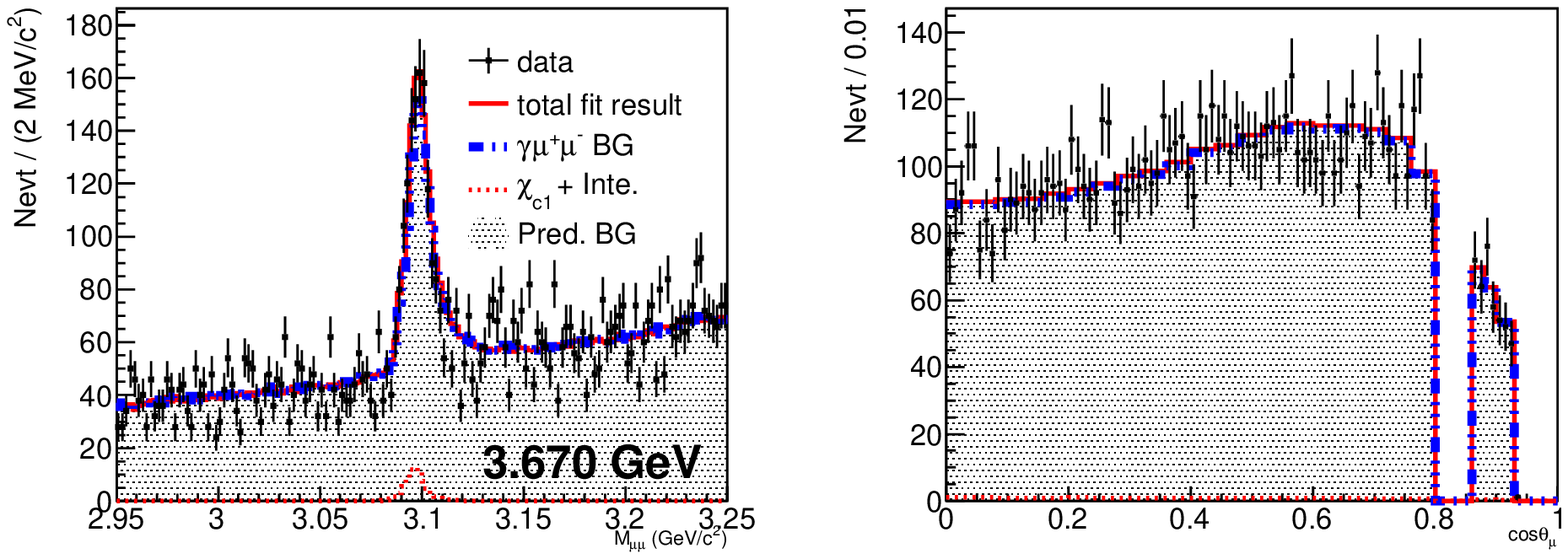}
\\
\end{center}
\caption{One-dimensional projections of the two-dimensional fit
to the $M_{\MM}$ and $|\cos\theta_{\mu}|$ distributions from the 
control data samples.
The two-dimensional correction is not applied in this fit.
The black dots with error bars are the distributions from data, 
the gray histograms
are the irreducible background predicted by the corrected MC simulation.
The red curve is the best fit result, 
the red dotted (blue dashed) curve is the signal (background)
contribution.
}
\label{fig:fit_control_sample_NoCorr}
\end{figure*}

\begin{figure*}[htb]
\begin{center}
\includegraphics[width=0.45\textwidth]
{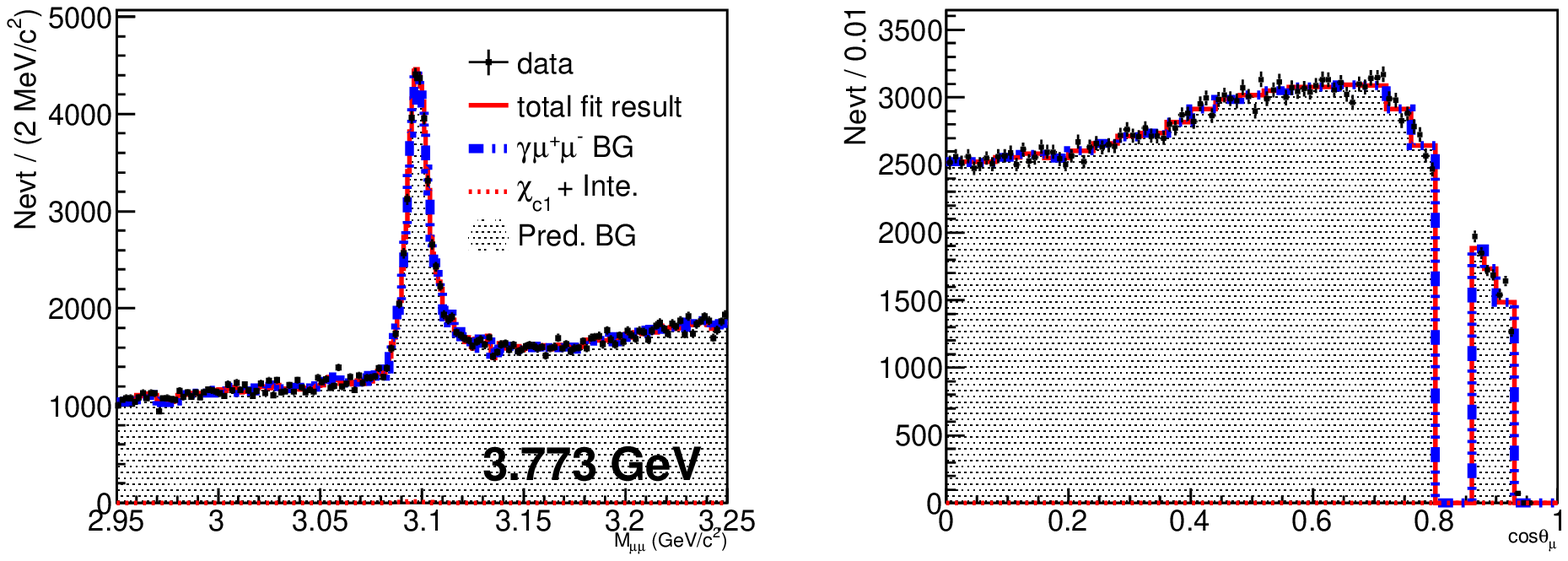}
\includegraphics[width=0.45\textwidth]
{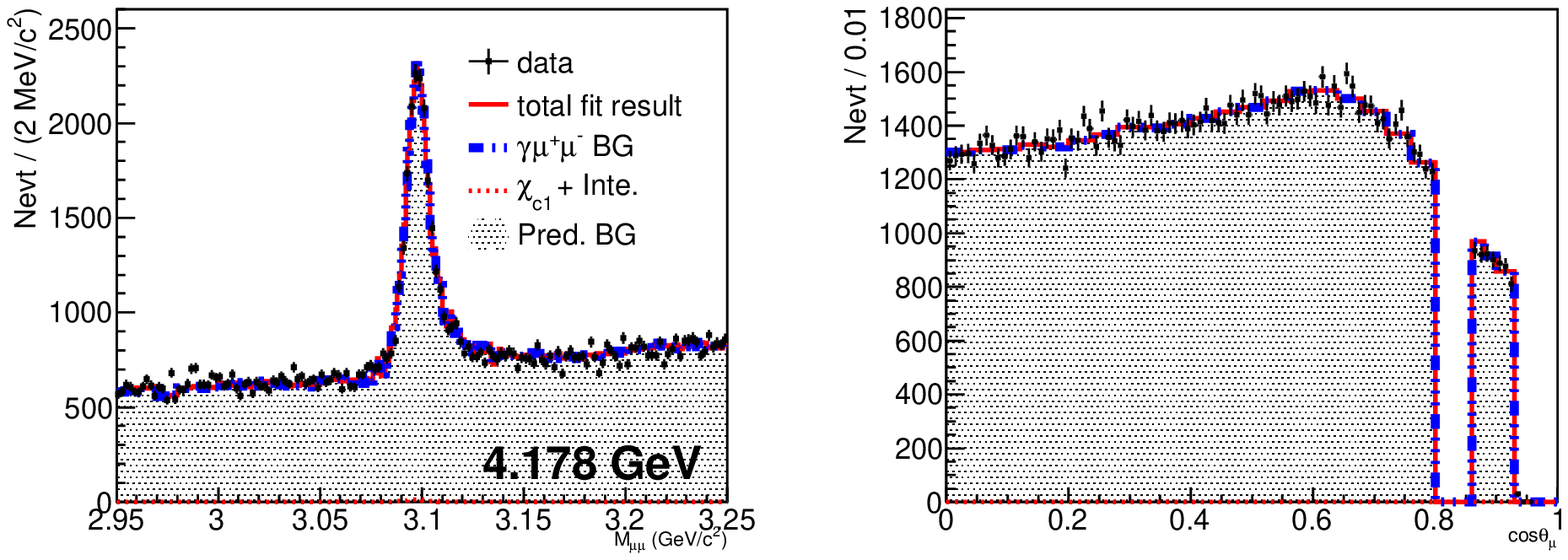}
\includegraphics[width=0.45\textwidth]
{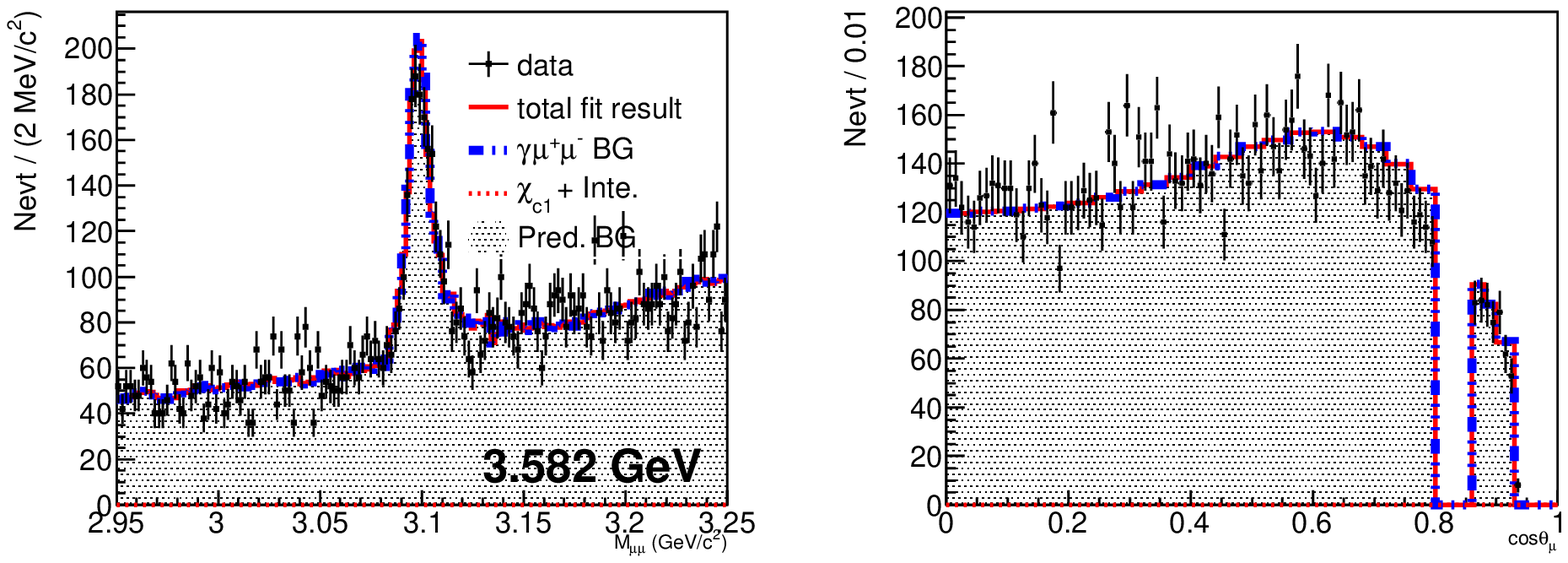}
\includegraphics[width=0.45\textwidth]
{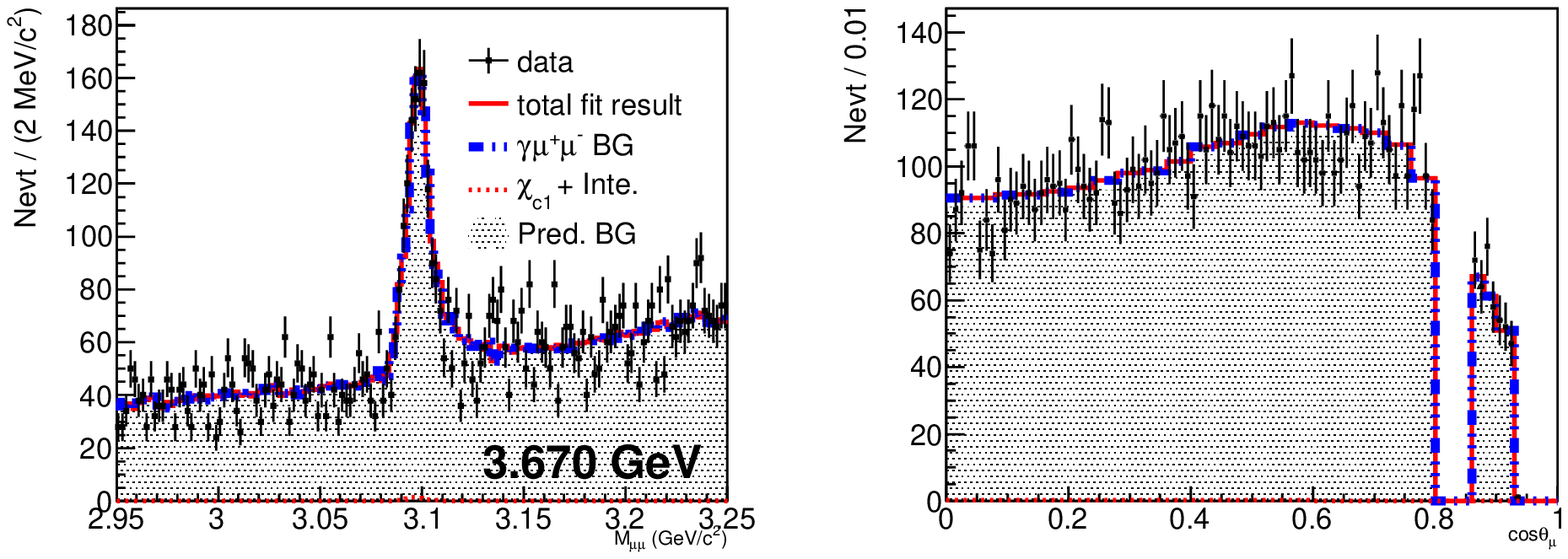}
\\
\end{center}
\caption{One-dimensional projections of the two-dimensional fit
to the $M_{\MM}$ and $|\cos\theta_{\mu}|$ distributions from the 
control data samples.
The two-dimensional correction factor is determined from 
$\sqrt{s}=3.773~\mathrm{GeV}$ sample.
}
\label{fig:fit_control_sample_3773Corr}
\end{figure*}

\begin{figure*}[htb]
\begin{center}
\includegraphics[width=0.45\textwidth]
{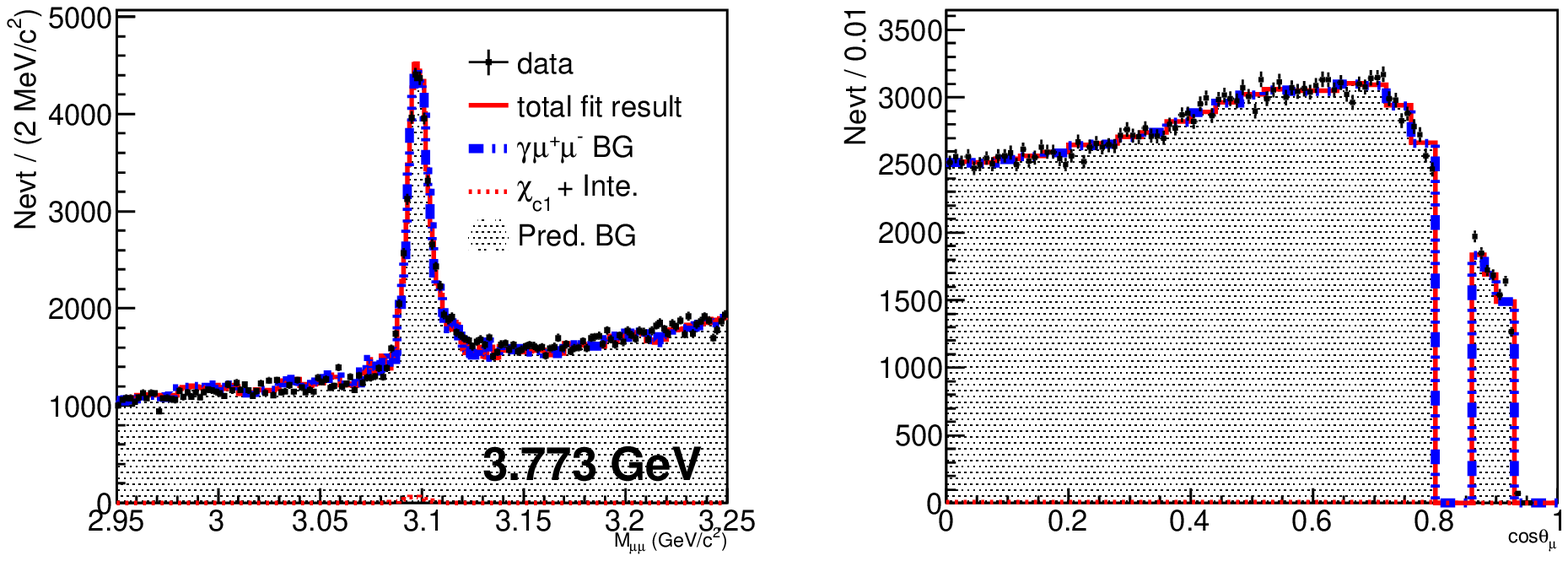}
\includegraphics[width=0.45\textwidth]
{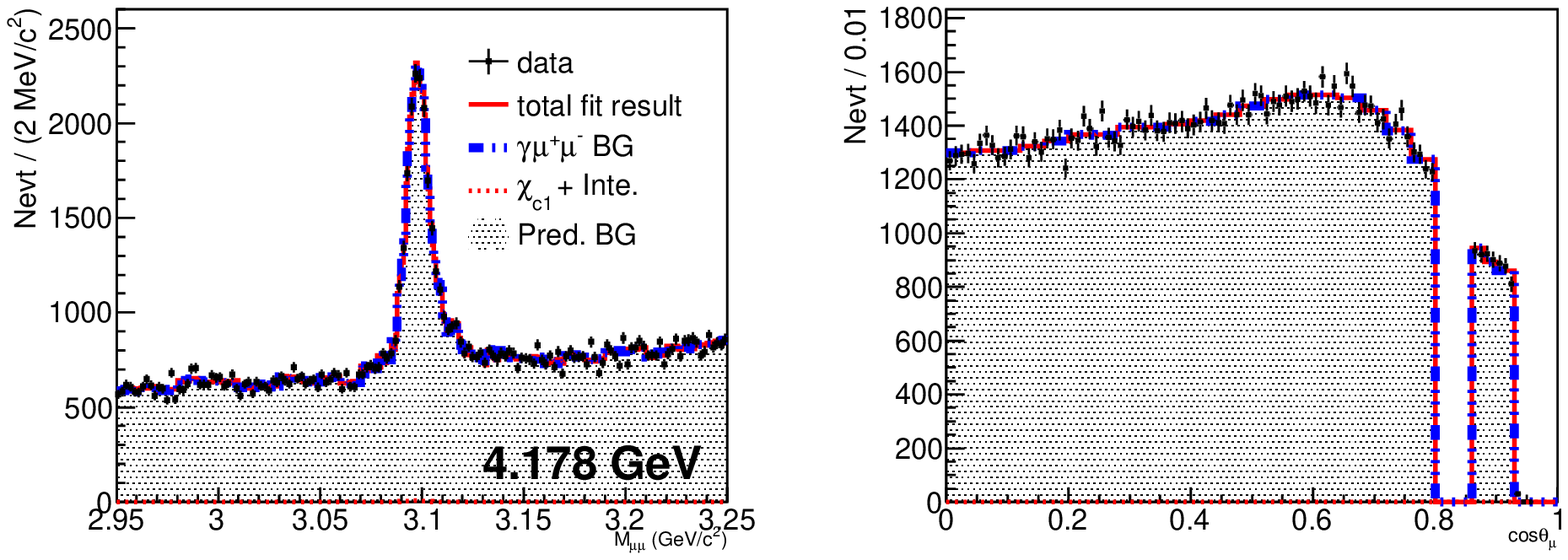}
\includegraphics[width=0.45\textwidth]
{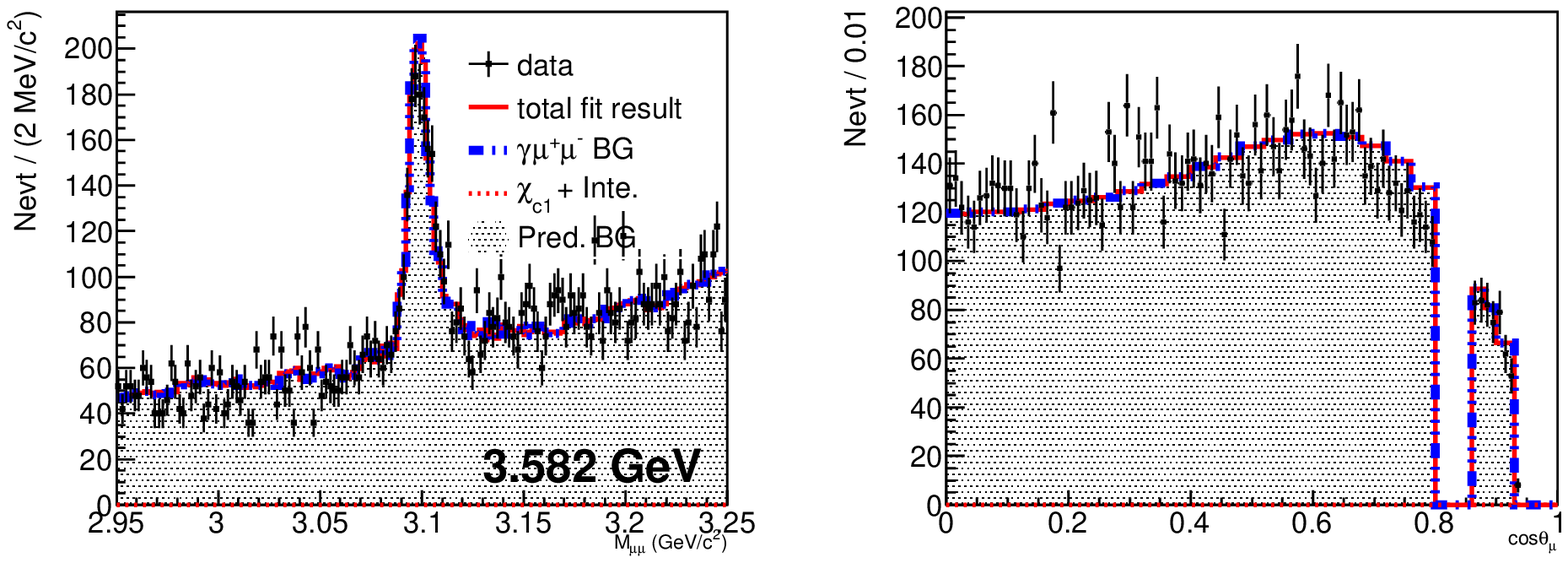}
\includegraphics[width=0.45\textwidth]
{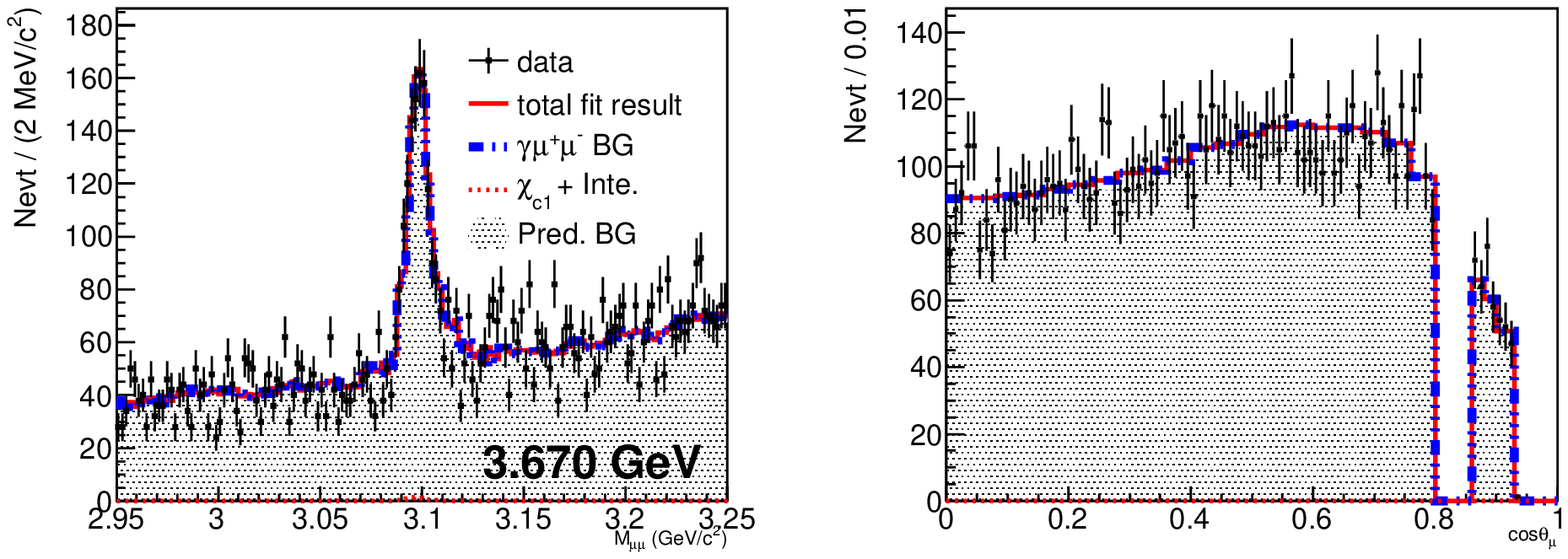}
\\
\end{center}
\caption{One-dimensional projections of the two-dimensional fit
to the $M_{\MM}$ and $|\cos\theta_{\mu}|$ distributions from the 
control data samples.
The two-dimensional correction factor is determined from 
$\sqrt{s}=4.178~\mathrm{GeV}$ sample.
}
\label{fig:fit_control_sample_4180Corr}
\end{figure*}

Figure~\ref{fig:corr_factor} shows the correction factors used for the two-dimensional
correction to the distribution of $M_{\mu^+\mu^-}$ and $|\mathrm{cos}\theta_\mu|$.
The left plot shows the correction factors derived from the $\sqrt{s}=3.773~\mathrm{GeV}$
sample and the right plot is from the $\sqrt{s}=4.178~\mathrm{GeV}$ sample. 
Figure~\ref{fig:fit_control_sample_NoCorr}, Fig.~\ref{fig:fit_control_sample_3773Corr},
and Fig.~\ref{fig:fit_control_sample_4180Corr} show the results from the two-dimensional 
fits to the $M_{\MM}$ and $|\mathrm{cos}\theta_{\mu}|$ distributions from the control
samples before correction, after correction using the correction factors extracted
from data and MC samples at $\sqrt{s}=3.773~\mathrm{GeV}$, and the correction factors
from $\sqrt{s}=4.178~\mathrm{GeV}$.

\section{The 2-dimensional fit method}

\begin{figure*}[htbp]
\begin{center}
\includegraphics[width=0.5\textwidth]
{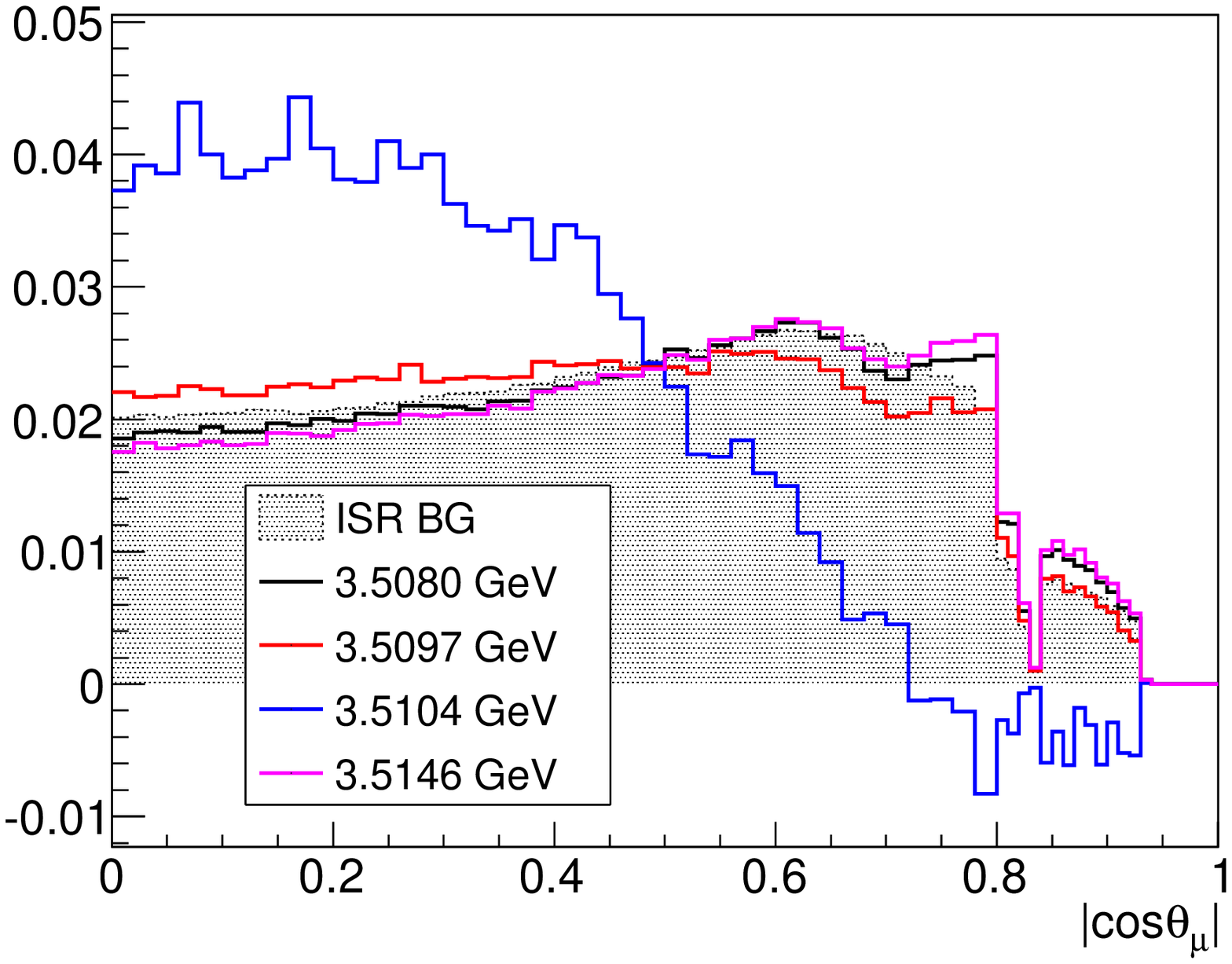}
\end{center}
\caption{Comparison of the line-shape of $|\mathrm{cos}\theta_{\mu}|$ from
the background simulation (green histogram)
and the signal MC simulation (other histograms).
}
\label{fig:compare cosmu}
\end{figure*}

Figure~\ref{fig:compare cosmu} shows the $|\mathrm{cos}\theta_{\mu}|$
distribution of the signal MC simulation at different center-of-mass energies,
compared with the distribution from the irreducible background MC simulation.
The signal MC samples are produced with $\Gamma_{ee}$ and $\phi$ fixed to
the best value determined from this study.

\section{scatter plot and chi distribution of $\cco$ scan samples}

\begin{figure*}[htbp]
\begin{center}
\includegraphics[width=0.85\textwidth]
{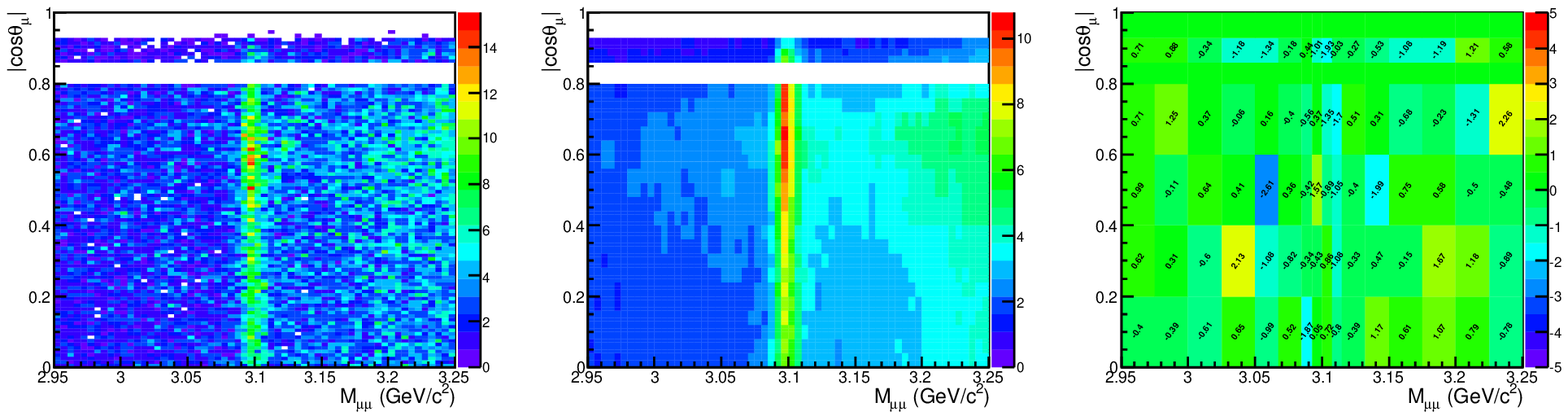}
\includegraphics[width=0.85\textwidth]
{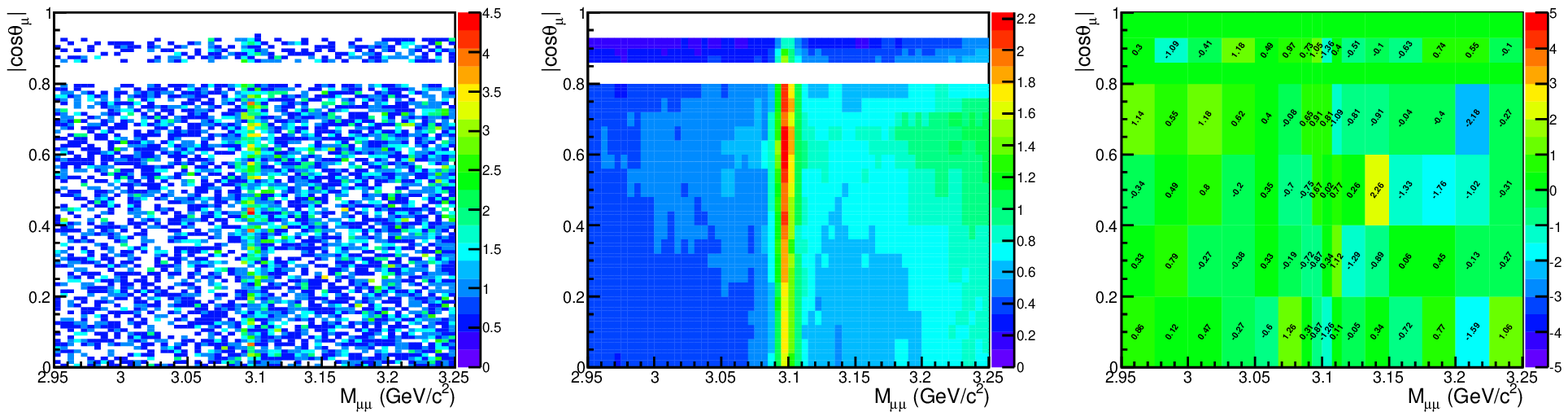}\\
\includegraphics[width=0.85\textwidth]
{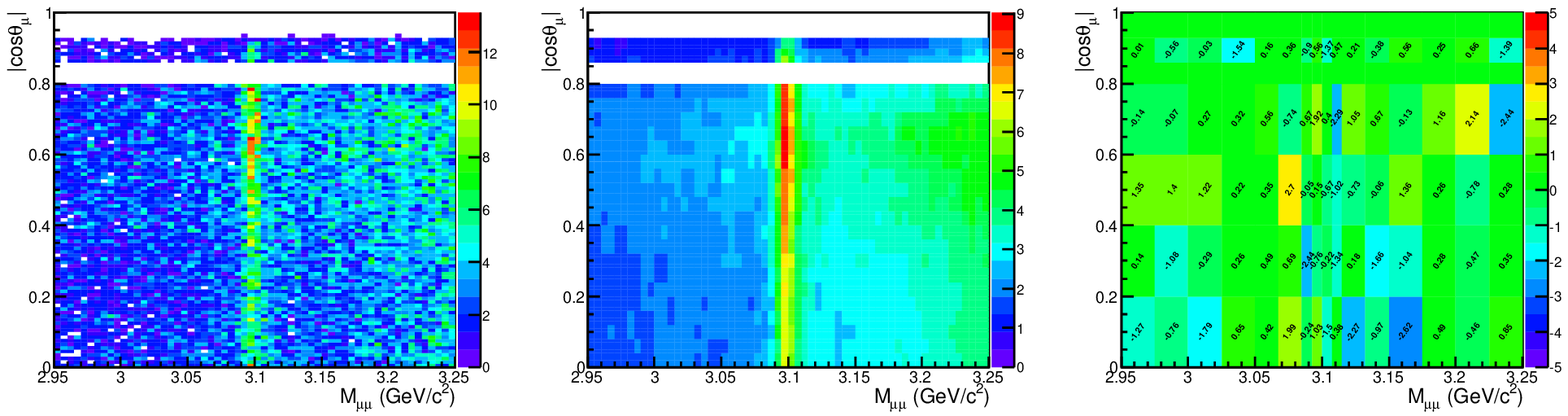}
\includegraphics[width=0.85\textwidth]
{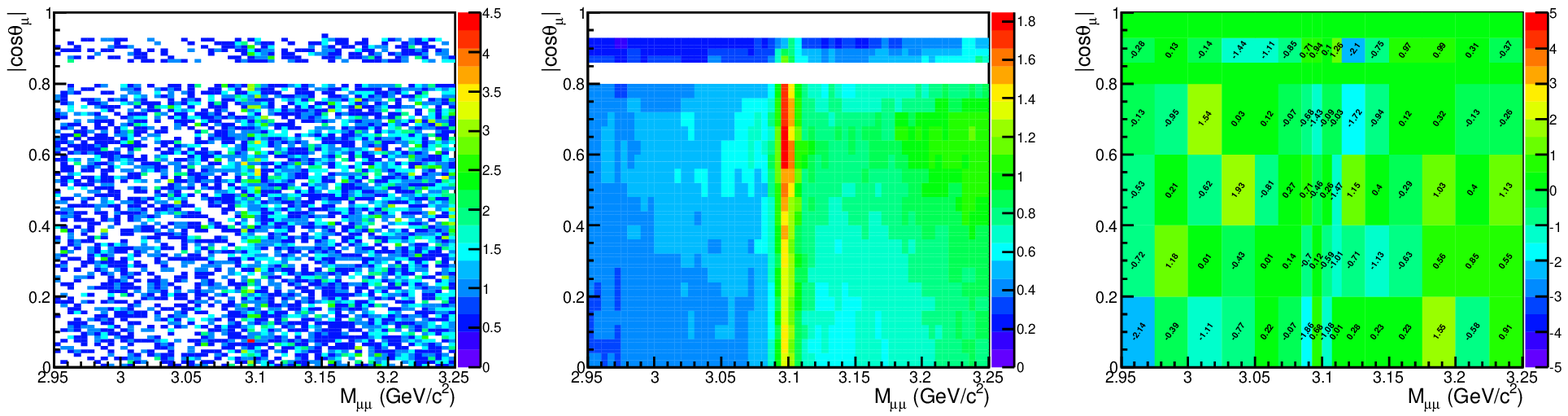}
\end{center}
\caption{
The scatter plots and pull distributions at 
$\sqrt{s}=3.5080$, $3.5097$,
$3.5104$, and $3.5146~\gev$.
}
\label{fig:chi distribution}
\end{figure*}

Figure~\ref{fig:chi distribution} shows the scatter plots of data (left), MC (middle), and the pull distributions from the two-dimentional fit (right) at $\cco$ scan samples.

\newpage

\section{Statistical test for the common fit}

\begin{figure*}[htbp]
\begin{center}
\includegraphics[width=0.4\textwidth]
{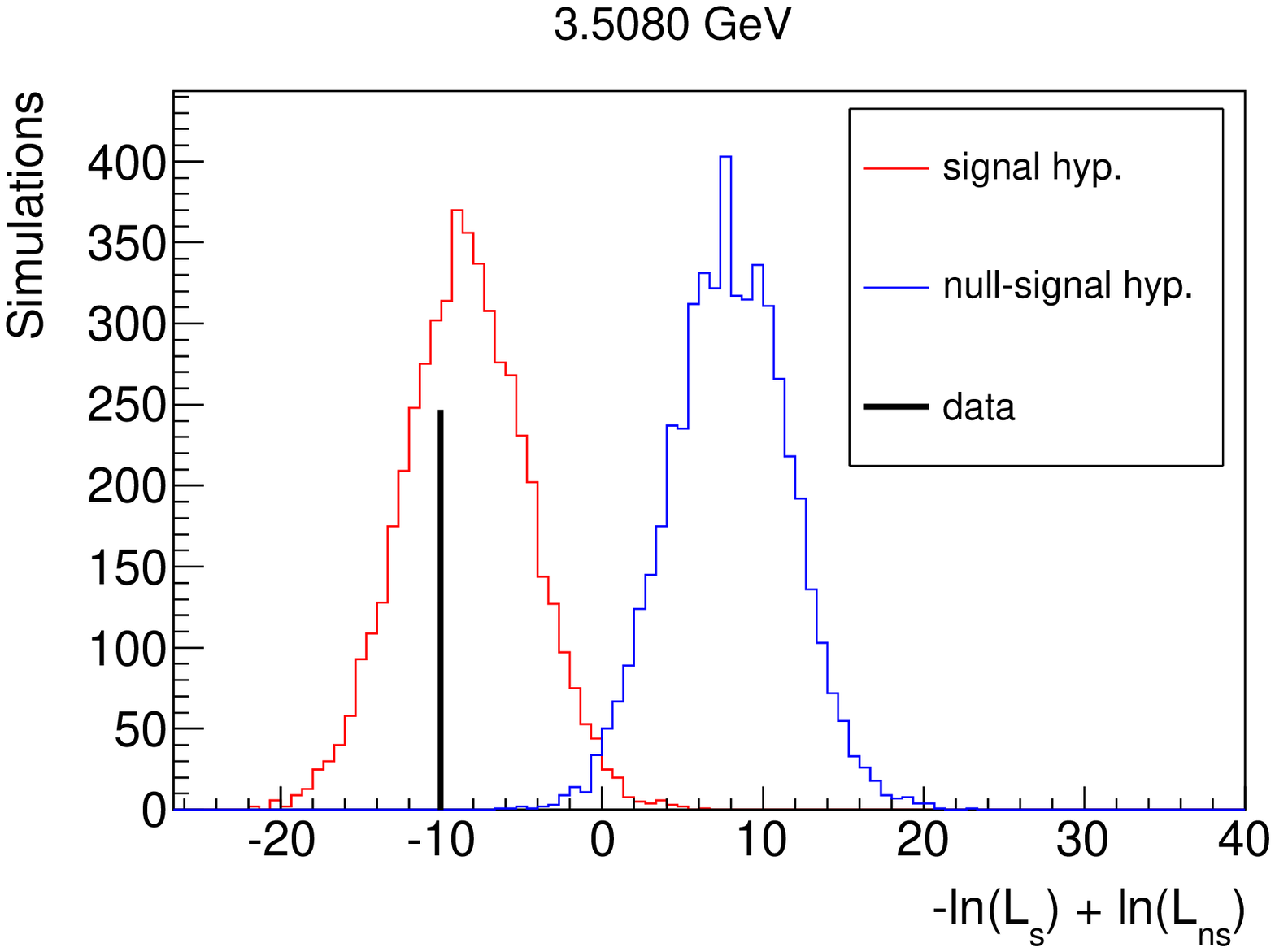}
\includegraphics[width=0.4\textwidth]
{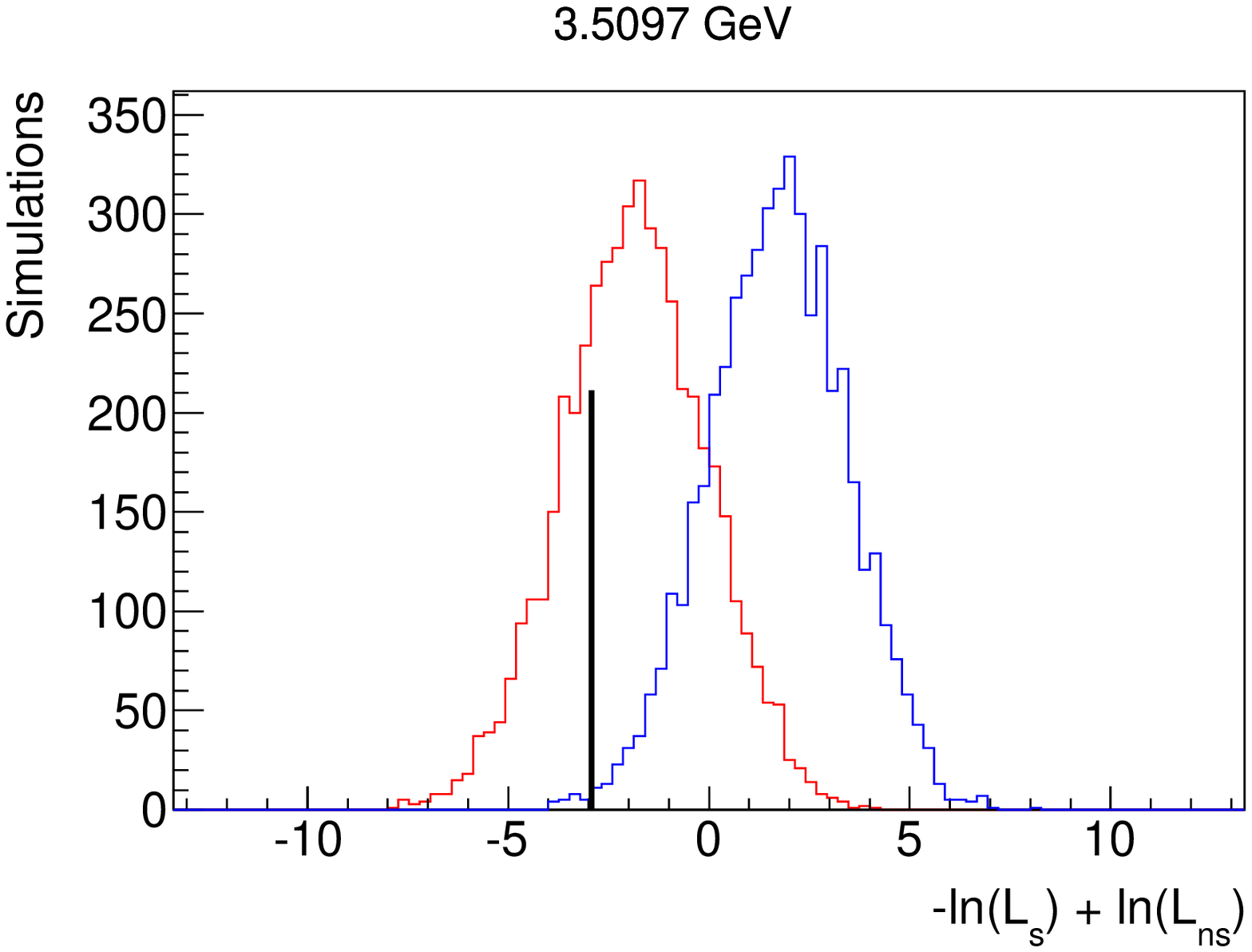}\\
\includegraphics[width=0.4\textwidth]
{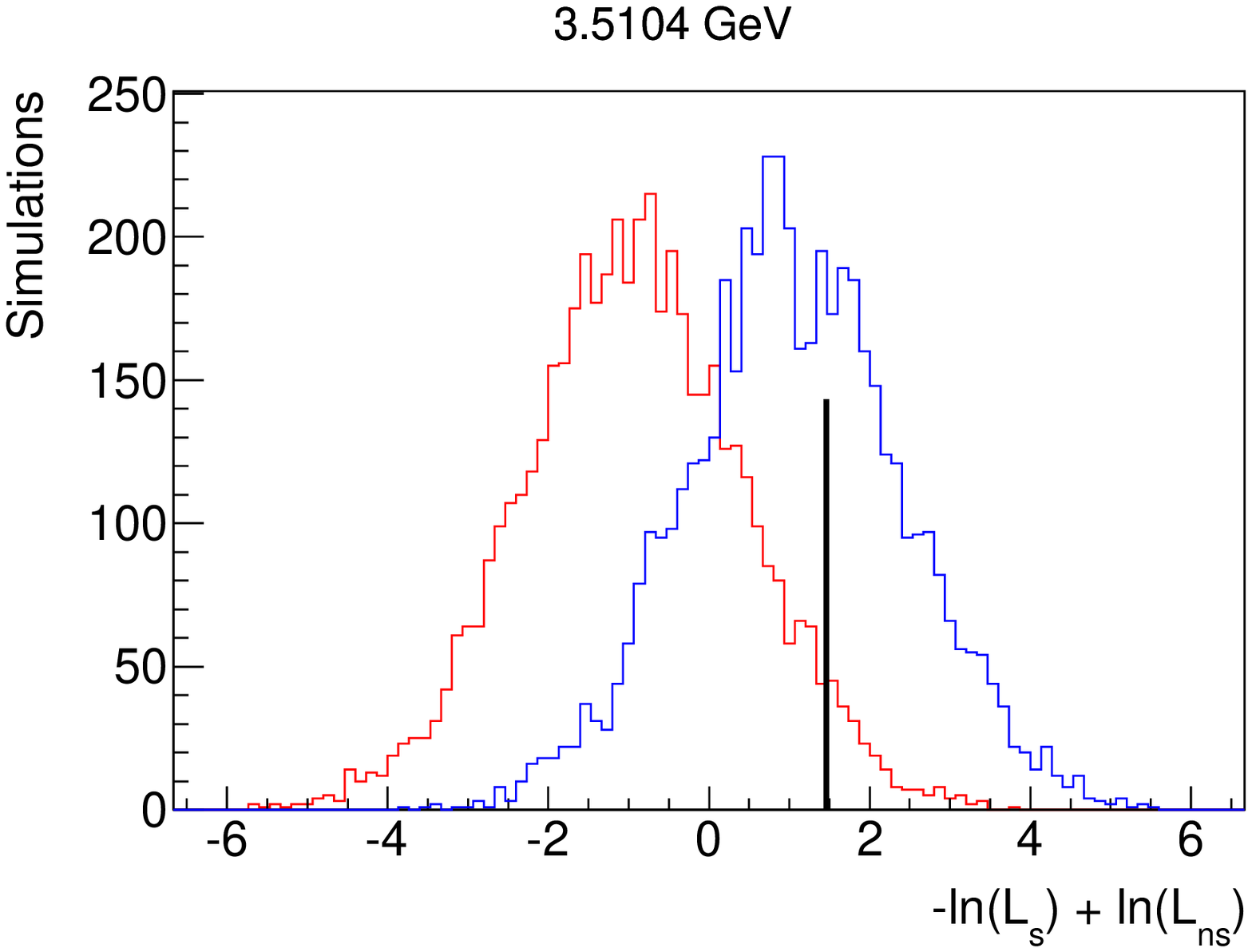}
\includegraphics[width=0.4\textwidth]
{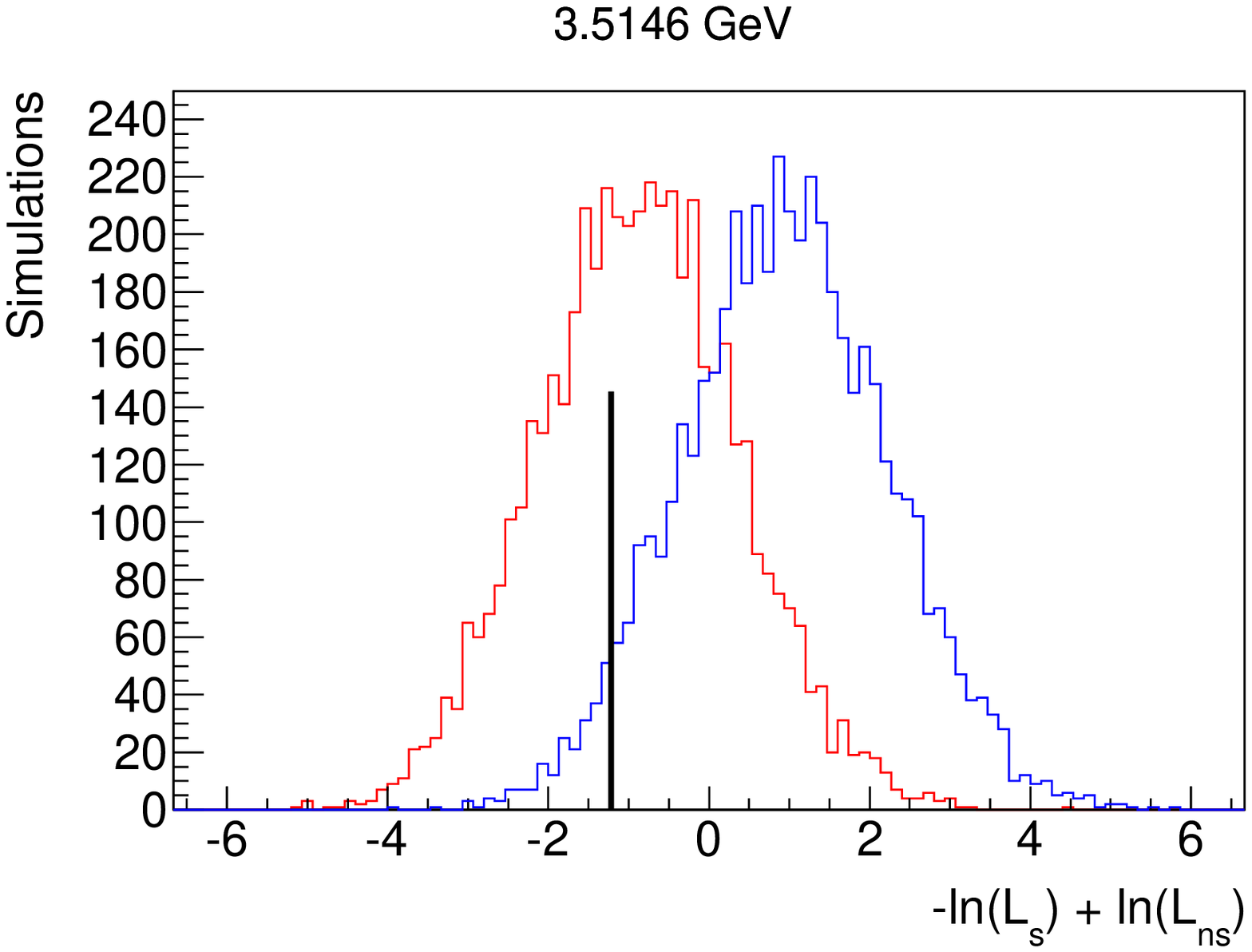}
\end{center}
\caption{Distributions of 
the test variable $t$ from the toy MC samples 
based on the common fit result under the signal and 
null-signal hypotheses at $\sqrt{s}=3.5080$, $3.5097$,
$3.5104$, and $3.5146~\gev$. The red and the blue histograms
show the distributions under the signal 
and null-signal hypotheses, respectively, 
while the black vertical lines indicate the values from real data.
}
\label{fig:hypothesises_likelihood}
\end{figure*}

\begin{figure*}[htbp]
\begin{center}
\includegraphics[width=0.4\textwidth]
{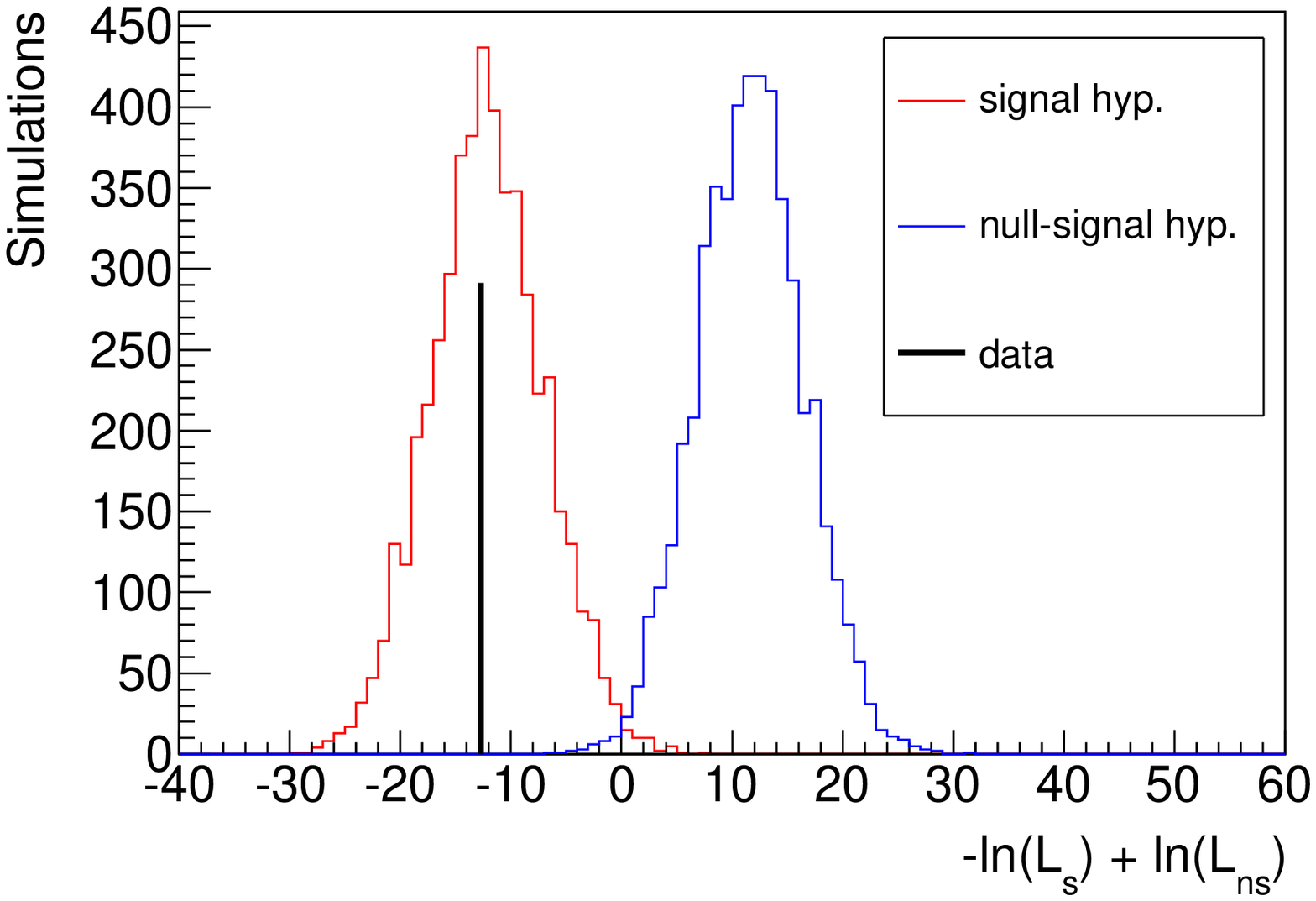}
\end{center}
\caption{Distribution of the test variable $t$ from the toy
MC samples using all four $\cco$ scan samples.
}
\label{fig:hypothesises_likelihood_tot}
\end{figure*}

For the $\cco$ scan samples,
statistical tests are performed by using the
toy MC samples based on the common fit result
under the signal and the null-signal hypotheses.
The difference of the log-likelihood values, 
$t=-\ln{L}_{s} + \ln{L}_{ns}$,
is used as a test variable,
where the signal hypothesis is given by $(-\ln{L}_{s})$
and the null-signal hypothesis by $(-\ln{L}_{ns})$.
The distributions of $t$ for the four $\cco$ scan samples
are shown in Fig.~\ref{fig:hypothesises_likelihood},
and the result combining the four samples
is shown in Fig.~\ref{fig:hypothesises_likelihood_tot}.

\newpage

